\documentclass[12pt,reqno]{amsart}
\usepackage{amssymb, amsmath, hyperref}
\usepackage{pdfsync}
\usepackage{stmaryrd}
\usepackage{bbold}
\usepackage{bbm}
\usepackage{graphics,graphicx,fancyhdr,color}

\newtheorem{df}{Definition}[section]
\newtheorem{lm}[df]{Lemma}
\newtheorem{lemma}[df]{Lemma}
\newtheorem{prop}[df]{Proposition}
\newtheorem{thm}[df]{Theorem}
\newtheorem{cor}[df]{Corollary}

\makeatletter \@addtoreset{equation}{section}

\newcommand{\cal}{\mathcal}

\newcommand{\bes}{\begin{displaymath}}
\newcommand{\ees}{\end{displaymath}}
\newcommand{\be}{\begin{equation}}
\newcommand{\ee}{\end{equation}}
\newcommand{\ba}{\begin{eqnarray}}
\newcommand{\ea}{\end{eqnarray}}
\newcommand{\bas}{\begin{eqnarray*}}
\newcommand{\eas}{\end{eqnarray*}}
\newcommand{\@Bbb}[1]{\ensuremath{\mathbb #1}}

\newcommand{\B}{{\@Bbb B}}
\newcommand{\C}{{\@Bbb C}}

\newcommand{\F}{{\@Bbb F}}
\renewcommand{\P}{{\mathbb P}}
\newcommand{\bbP}{{\P}}
\newcommand{\bbE}{{\mathbb E}}
\newcommand{\Q}{{\@Bbb Q}}
\newcommand{\bQ}{{\@Bbb Q}}
\newcommand{\N}{{\@Bbb N}}
\newcommand{\R}{{\@Bbb R}}

\newcommand{\bbR}{{\@Bbb R}}
\newcommand{\W}{{\@Bbb W}}
\newcommand{\Z}{{\@Bbb Z}}
\newcommand{\bbZ}{{\@Bbb Z}}
\newcommand{\bbT}{{\@Bbb T}}

\newcommand{\la}{\lambda}
\newcommand{\al}{\alpha}

\newcommand{\si}{\sigma}

\newcommand{\Om}{\Omega}
\newcommand{\om}{\omega}

\newcommand{\ep}{\varepsilon}
\newcommand{\eps}{\epsilon}

\newcommand{\@s}[1]{\ensuremath{\mathcal #1}}
\newcommand{\cA}{\@s A}
\newcommand{\cB}{\@s B}
\newcommand{\cC}{\@s C}
\newcommand{\cD}{\@s D}
\newcommand{\cE}{\@s E}
\newcommand{\cF}{\@s F}
\newcommand{\cG}{\@s G}
\newcommand{\cH}{\@s H}
\newcommand{\cI}{\@s I}
\newcommand{\cJ}{\@s J}

\newcommand{\cK}{\@s K}
\newcommand{\cL}{\@s L}
\newcommand{\cN}{\@s N}
\newcommand{\cM}{\@s M}
\newcommand{\cO}{\@s O}
\newcommand{\cP}{\@s P}
\newcommand{\cR}{\@s R}
\newcommand{\cS}{\@s S}
\newcommand{\cT}{\@s T}
\newcommand{\cV}{\@s V}
\newcommand{\cW}{\@s W}
\newcommand{\cX}{\@s X}
\newcommand{\cY}{\@s Y}
\newcommand{\cZ}{\@s Z}

\newcommand{\@bm}[1]{\ensuremath{\mathbf #1}}
\newcommand{\bma}{\@bm a}
\newcommand{\bmb}{\@bm b}
\newcommand{\bmc}{\@bm c}
\newcommand{\bmd}{\@bm d}

\newcommand{\bme}{\@bm e}
\newcommand{\bmf}{\@bm f}
\newcommand{\bmg}{\@bm g}
\newcommand{\bmh}{\@bm h}
\newcommand{\bmi}{\@bm i}
\newcommand{\bmj}{\@bm j}
\newcommand{\bmk}{\@bm k}
\newcommand{\bml}{\@bm l}
\newcommand{\bmm}{\@bm m}
\newcommand{\bmn}{\@bm n}
\newcommand{\bmo}{\@bm o}
\newcommand{\bmp}{\@bm p}
\newcommand{\bmq}{\@bm q}
\newcommand{\bmr}{\@bm r}
\newcommand{\bms}{\@bm s}
\newcommand{\bmt}{\@bm t}
\newcommand{\bmu}{\@bm u}
\newcommand{\bmw}{\@bm w}
\newcommand{\bmv}{\@bm v}
\newcommand{\bmx}{\@bm x}
\newcommand{\bx}{\@bm x}

\newcommand{\bmy}{\@bm y}
\newcommand{\bz}{\@bm z}

\newcommand{\by}{\@bm y}

\newcommand{\bmzero}{\@bm 0}

\newcommand{\ga}{\gamma}

\newcommand{\@g}[1]{\ensuremath{\mathfrak #1}}
\newcommand{\gA}{\@g A}
\newcommand{\gD}{\@g D}
\newcommand{\gJ}{\@g J}
\newcommand{\gF}{\@g F}
\newcommand{\gM}{\@g M}
\newcommand{\gR}{\@g R}

\newcommand{\gq}{\@g q}
\newcommand{\gr}{\@g r}
\newcommand{\gp}{\@g p}
\newcommand{\gen}{\@g e}

\newcommand{\blambda}{\boldsymbol\lambda}

\newcommand{\commentout}[1]{{}}

\makeatother

\title[Ballistic and Superdiffusive Transport]
{Ballistic and superdiffusive scales in macroscopic evolution of a
  chain of oscillators}

\begin{document}

\author{Tomasz Komorowski}
\address{Tomasz Komorowski\\Institute of Mathematics, Polish Academy Of Sciences\\Warsaw, Poland.}
\email{{\tt komorow@hektor.umcs.lublin.pl}}
\author{Stefano Olla}
\address{Stefano Olla\\
 CEREMADE, UMR-CNRS 7534\\
 Universit\'{e} Paris Dauphine\\
 Paris, France.}
 \email{{\tt olla@ceremade.dauphine.fr}}

 \begin{abstract}
{\em We consider a one dimensional infinite acoustic chain of harmonic
  oscillators whose 
  dynamics is perturbed by  a random exchange of velocities, such that the energy and
  momentum of the chain are conserved. Consequently, the evolution of the system has only three
  conserved quantities: volume, momentum and energy. 
  We show the existence of two space--time scales on which the energy of
  the system evolves. On the hyperbolic scale
  $(t\eps^{-1},x\eps^{-1})$ the limits of the conserved quantities
  satisfy a Euler system of equations, 
  while the thermal part of the energy macroscopic profile
  remains stationary. 
   Thermal energy
  starts evolving at a longer time scale, corresponding to the superdiffusive scaling
  $(t\eps^{-3/2},x\eps^{-1})$ and 
  follows a fractional heat equation. We also prove the diffusive
  scaling limit of the Riemann invariants - the so
  called normal modes, corresponding to the linear hyperbolic propagation
  .}
\end{abstract}

\thanks{This work has been partially supported by the
  European Advanced Grant {\em Macroscopic Laws and Dynamical Systems}
  (MALADY) (ERC AdG 246953), and by the CAPES
    and CNPq program \emph{Science Without Borders}. Work of T.K. has
    been partially supported by the Polish National
    Science Centre grant UMO-2012/07/B/SR1/03320.}

\date{\today. {\bf File: {\jobname}.tex.}}

\maketitle

\section{Introduction}


Consider a chain
of coupled anharmonic oscillators in one 
dimension. 
Denote by ${\frak q}_x$ and $\frak p_x$ the position and momentum of
the particle labeled by $x\in\mathbb Z$.  
The interaction between  particles $x$ and $x-1$ is described by
the potential energy $V(\frak r_x)$ of an anharmonic spring, where
 the quantity
 \begin{equation}
\label{strain}
{\frak r}_x:={\frak q}_x-{\frak q}_{x-1}
\end{equation}
is
 called the inter-particle
 distance or \emph{volume strain}. 
The (formal) Hamiltonian of the chain is given by 
\begin{equation}
\label{hamilton}
{\cal H}({\frak
  q},{\frak p})=\sum_x{\frak e}_x ({\frak
  q},{\frak p}),
\end{equation}
 where the energy of the oscillator
$x$ is defined by
\begin{equation}
\label{011705}
\frak e_x ({\frak q}, {\frak p}):= \frac{\gp_x^2}{2}+V(\gr_x) .
\end{equation}
The respective Hamiltonian dynamics is given by the solution of the equations:
  \begin{eqnarray} 
&&\dot {\frak q}_{x}(t)= \frac{\partial {\cal H}}{\partial\frak p_x} ({\frak
  q},{\frak p})=\frak p_x,
\label{eq:bash}\\
&&\nonumber\\
&& \dot {\frak p}_x(t)=-\frac{\partial {\cal H}}{\partial\frak q_x} ({\frak
  q},{\frak p}) 
   , 
\quad x\in\bbZ.\nonumber 
\end{eqnarray}
There are three formally conserved quantities (also called balanced)
for this dynamics
\begin{equation}
  \label{eq:2}
  \sum_x \gr_x, \qquad \sum_x \gp_x, \qquad \sum_x \gen_x
\end{equation}
that correspond to the total volume, momentum and energy of the chain. 
The corresponding equilibrium Gibbs measures $\nu_{\blambda}$ are parametrized by 
$\blambda= [\beta, p,\tau ]$, with the respective components
equal to the inverse temperature, velocity and tension.  
They are product measures  given explicitly by formulas
\begin{equation}
  \label{eq:gibbs}
  d\nu_{\blambda} = \prod_x \exp\left\{-\beta \left(\vphantom{\int_0^1}\gen_x - p \gp_x
    -\tau \gr_x\right)- \mathcal G(\blambda)\right\} d\gr_x \; d\gp_x, 
\end{equation}
where
\begin{equation}
\label{011911}
\mathcal
G(\blambda):=\beta\frac{p^2}{2}+\log\left[\sqrt{2\pi\beta^{-1}}
Z( \tau,\beta)\right] 
\end{equation}
and
\begin{equation}
\label{011911a}
Z(\tau ,\beta):=\int_{\bbR}\exp\left\{-\beta \left(V(r)
    -\tau r\right) \right\} dr.
\end{equation}
The length and  internal energy in equilibrium can be expressed as
functions of $\blambda$ by
\begin{eqnarray}
\label{tau-n}
&&
 r(\tau,\beta)=\langle {\frak r}_x\rangle_{\nu_{\blambda}}
= \beta^{-1} \partial_\tau  \mathcal G(\blambda), \qquad
p=\beta^{-1}\partial_p  \mathcal G(\blambda)
\\
&&
\nonumber\\
&& u(\tau,\beta)=\langle {\frak e}_x \rangle_{\nu_{\blambda}}-\frac{p^2}{2} =\frac{1}{2\beta}+\langle V({\frak r}_x) \rangle_{\nu_{\blambda}}=
   -\partial_\beta  \mathcal G(\blambda)+\tau r. 
   \nonumber
\end{eqnarray}
Therefore the tension $\tau(r, u)$ and
inverse temperature $\beta(r, u)$  can be determined
from the relations:
\begin{equation}
  \label{eq:inv}
  \beta = \partial_u S(u,r) , \qquad \tau\beta = - \partial_r S(u,r),
\end{equation}
where $S(u,r)$ is {\em the thermodynamic entropy} defined by the Legendre
transform
\begin{equation}
  \label{eq:3}
  S(u,r) = \inf_{\tau,\beta} \left[ -\beta\tau r + \beta u +
 \mathcal     G(\beta,0,\tau) \right] .
\end{equation}
It is expected that, after hyperbolic rescaling of space and time $(t\eps^{-1},x\eps^{-1})$,  the
empirical distribution of the balanced quantities  
\begin{equation}
\label{w}
\frak w_x^T(t) :=
[{\frak r}_x(t), {\frak p}_x(t),{\frak e}_x (t)],
\end{equation} defined for a smooth
function $J$ with compact support by the formula:
\begin{equation*}
 \eps \sum_x J(\eps x)   \frak w_x\left(\frac{t}{\eps} \right)
\end{equation*}
converges, as $\eps\to 0$, to the solution $w^T(t,y) = \left[r(t,y),
  p(t,y), e(t,y)\right]$ of {\em the compressible Euler system} of equations:
\begin{equation}
  \label{eq:euler}
  \begin{split}
    \partial_t r &= \partial_y p,\\
    \partial_t p &= \partial_y \tau( r, u),\\
    \partial_t e &= \partial_y \left[ p \tau( r, u)\right],\\ 
      \end{split}
\end{equation}
where $u (t,y):= e (t,y)- p^2(t,y)/2$ is the local internal energy.
For a finite macroscopic volume, this limit has been proven using
the relative entropy method, provided the microscopic dynamics is
perturbed by a random exchange of velocities between particles, see
\cite{OVY} and \cite{EO},  in the regime when  the system of Euler
equations admits a smooth solution.
After a sufficiently long time, the solution of \eqref{eq:euler}
should converge, in an appropriately weak sense, to some mechanical equilibrium described  by:
\begin{equation}
  \label{eq:mechequi}
   p(x) = p_0, \qquad \tau( r(x), u(x)) = \tau_0,
\end{equation}
 where $p_0$ and $\tau_0$ are some constants.

To characterize all possible stationary
solutions of \eqref{eq:mechequi} can possibly be a daunting task. Most likely  they are generically very
irregular. But it is quite obvious that if we start with smooth
initial conditions 
that satisfy \eqref{eq:mechequi}, the respective solutions of
\eqref{eq:euler} remain stationary. Also by the same
relative entropy method as used in \cite{OVY} and \cite{EO}, it
follows that starting with
such initial profiles, the corresponding empirical distributions of
the balanced quantities converge, in the hyperbolic time scale, to the 
same initial stationary solution, i.e. they do not evolve in time. 
On the other hand, we do know that the system will eventually converge
to a global equilibrium, 
so this implies that there exists a longer time scale, on which
these profiles (stationary at the hyperbolic scale) will evolve. 

There is a strong argument, stemming from both a numerical evidence
and quite convincing heuristics,  
suggesting  the divergence of the Green-Kubo formula, defining the
thermal diffusivity, for a generic unpinned one dimensional system,
see \cite{sll} \cite{llp97}. 
Therefore, we expect that the aforementioned larger time scale (at which 
the evolution of these profiles is observed) is superdiffusive.  
Furthermore, an argument  by H. Spohn \cite{Sp13}, based
  on fluctuating hydrodynamics and mode coupling, also suggests a
  superdiffusive evolution of the \emph{heat mode} fluctuation field,
  when the system is in equilibrium. Consequently, one can conjecture
the following scenario: after the space-time rescaling 
 $(\eps^{-\alpha} t,\eps^{-1}x)$,  the temperature $T(t,y) = \beta^{-1}(t,y)$
evolves following some fractional (possibly non-linear) heat equation. 
The choice of the exponent $\alpha$ may depend on the particular values of the
tension and of the interaction potential $V(r)$.

In the present work we prove rigorously that this picture of two time scale
evolution holds for the harmonic
chain with a momentum exchange noise that conserves the total momentum
and energy. 
 We consider the quadratic Hamiltonian: 
 \begin{equation}
   \label{eq:hquad}
   {\cal H}({\frak q},{\frak p}) = \sum_x 
   \frac12{\frak p}_x^2 -\frac{1}{4}
     \sum_{x,x'}\alpha_{x-x'}({\frak q}_x-{\frak q}_{x'})^2
 \end{equation}
where  harmonic coupling coefficients $\alpha_{x-x'}$, between atoms labelled by $x$ and $x'$, are assumed to  decay sufficiently fast. 
The  detailed hypotheses made about the potential
are contained in Section \ref{sec-bas}.
Furthermore we perturb the Hamiltonian dynamics, given by
\eqref{eq:bash}, by a random exchange of momentum between the nearest
neighbor atoms, in such a way that the total energy and  momentum of the chain are
conserved,
see formula \eqref{eq:bas2} below. 
We assume that the initial configuration is distributed according to a 
probability measure
$\mu_\eps$  such that its total energy grows like $\eps^{-1}$ (i.e. macroscopically, it is of order $1$). 
Additionally, we suppose that the family of measures $\left(\mu_\eps\right)_{\eps>0}$
possesses a {\em macroscopic profile} $ w^T(y)=[ r(y),  p(y),  e(y)]$,
  with $ r, p, e$ belonging to $C_0^{\infty}(\bbR)$ -  the space of all smooth and compactly supported functions. The above means that
\begin{equation}
\label{010702x}
\lim_{\eps\to0+}\eps\sum_x\langle {\frak w}_x \rangle_{\mu_{\eps}}
J(\eps x) =\int_{\bbR} w^T(y) J(y)dy,
\end{equation}
for any test function $J\in C_0^\infty(\bbR)$.

We decompose the initial configuration into a phononic (low frequencies)
and thermal (high frequences) terms, i.e. 
the initial configuration $( {\frak  r}, {\frak p})$
is assumed to be of the form 
\begin{equation}
\label{conf}
{\frak  r}_x= {\frak  r}_x'+{\frak  r}_x''\quad \mbox{ and } 
\quad {\frak p}_x={\frak p}'_x+{\frak p}_x'',\quad x\in\bbZ,
\end{equation} 
where  the variance of
$( {\frak  r}_x'', {\frak p}_x'')$ around the macroscopic profile
$( r(\eps x), p(\eps x))_{x\in\bbZ}$ vanishes, 
with $\eps\to0+$ (see \eqref{010702} and
\eqref{010702a} below). This obviously implies that 
 the macroscopic profiles corresponding to   $( {\frak  r}_x', {\frak
   p}_x')$ equal to $0$. 
Concerning the distribution of this configuration we  suppose  that
its energy spectrum, defined in
\eqref{022403} below, is square integrable in the sense of condition
\eqref{finite-energy1}.
We call the respective energy profiles corresponding to  
$ ({\frak  r}_x', {\frak p}_x')$ and
$ ({\frak  r}_x'', {\frak p}_x'')$ as the 
thermal and phononic (or mechanical)  ones, see Definitions \ref{thermal} and
\ref{phononic} below.
{In Section \ref{sec10} we give several examples of
  families of 
  initial configurations satisfying the above hypotheses. Among them are inhomogeneous Gibbs measures with the 
  temperature, momentum and tension profiles changing on the
  macroscopic scale, see Section \ref{sec:local-gibbs-measures}. These
  examples include also 
  the families of measures corresponding to the local
  equilibrium states for the dynamics described by \eqref{hamilton},
  see Section \ref{lgg}.}

In our first result, see Theorem \ref{cor011002} and Corollary
\ref{cor031102} below, 
we show that  the 
empirical distribution of 
$\frak w_x (t)
$ converges, at the hyperbolic time-space scale, to  a solution
satisfy the linear version of \eqref{eq:euler} with
$\tau( r, u):=\tau_1 r$, where the parameter $\tau_1$,  called {\em the  speed of sound}, is given by formula
\eqref{tau}.

Notice that since we are working in infinite volume, relative entropy
methods cannot be applied to obtain such hydrodynamic limit, as done in \cite{EO} and \cite{OVY}. 
Instead, we obtain the limit result, using only the techniques based on the $L^2$
bounds on the initial configurations.
After a direct computation it becomes clear that the function
\begin{equation}
\label{Ty}
T(y):= e(t,y)- e_{\rm ph}(t,y),
\end{equation}
is stationary in time. 
 {\em The macroscopic phononic energy} is given by
\begin{equation}
\label{eph}
 e_{\rm ph}(t,y):= \frac{ p^2(t,y)}{2}+\tau_1 r^2(t,y),
\end{equation}
Also,  the entropy in this case  equals
$$
S(u,r)=C\log(u-\tau_1 r^2)
$$
for some constant $C>0$ depending on coefficients $(\al_x)$.
In fact, as it turns out, see Theorem
\ref{cor011002} below,
 $T(y)$ and $e_{\rm
  ph}(t,y)$, given by \eqref{Ty} and \eqref{eph} above,  are the respective limits of the thermal and  phononic  energy profiles.

Concerning the behavior of the thermal energy profile at the longer, 
superdiffusive time scale it has been shown in \cite{JKO} that after
the space-time rescaling  $(\eps^{-3/2} t,\eps^{-1}x)$,  the thermal
energy profile converges to the solution $T(t,y)$ 
of a fractional heat equation 
\begin{equation}
\label{frac-eqt}
\partial_t
T(t,y)=-\hat c|\Delta_y|^{3/4}T(t,y),\quad \mbox{ with }T(0,y)=T(y)
\end{equation} 
and the coefficient $\hat c$ given by formula \eqref{hatc-32a} below. This result
is generalized in the present paper to configurations of the form
\eqref{conf}. 

Finally, in Theorem \ref{p-modes} we consider the empirical distributions
of the microscopic estimators of the Riemann invariants of the linear
wave equation system that describe the evolution of the macroscopic profiles
of $({\frak r}_x(t),{\frak p}_x(t))$ - the so called {\em normal
  modes}. We show that they evolve at the
diffusive space-time scale $( t\eps^{-2},y\eps^{-1})$.
A similar behavior is conjectured for some an-harmonic chains,
e.g. those corresponding to the $\beta$-FPU potential at zero tension (cf. \cite{Sp13}).

Concerning the organization of the paper in Section \ref{sec2.0} we
rigorously introduce the basic notions that appear throughout the
article. The main results are formulated in Section \ref{sec3a}. In
Sections \ref{sec4.0} -- \ref{sec9} we present their respective
proofs. Finally in Section \ref{sec10} we show examples of
distributions of the initial data that are both of phononic and
thermal types. 


\section{The stochastic dynamics}

\label{sec2.0}


\subsection{Hamiltonian dynamics with a noise}

\label{sec-bas}

Concerning the interactions appearing in \eqref{eq:hquad} we consider
only the unpinned case, therefore we let 
\begin{equation}\label{al-0}
\alpha_0 := - \sum_{x\neq 0}
\alpha_x.
\end{equation} Define the Fourier transform of  $(\alpha_x)_{x\in\bbZ}$
by 
  \begin{equation}
\label{om2b}
\hat \al(k):=\sum_x\al_x\exp\left\{-2\pi i kx\right\},\quad k\in\bbT,
\end{equation}
where $\bbT$ is the unit torus, that is the  interval $[-1/2,1/2]$ with the identified endpoints. Then, from \eqref{al-0}, we have  $\hat\alpha(0) = 0$.

Furthermore, we assume  that:
 \begin{itemize}
 \item[a1)] coefficients  $(\alpha_x)_{x\in\bbZ}$ are real valued,
   symmetric and there exists a constant 
   $C>0$ such that 
   $$
   |\alpha_x|\le Ce^{-|x|/C}\quad\mbox{ for all }x\in \bbZ.
   $$ 
  \item[a2)] stability: $\hat\al(k)>0$, $k\not=0$,
\item[a3)] the chain is acoustic : $\hat \alpha''(0) > 0$,
 
 \end{itemize}


    The above conditions imply that 
   $\hat\alpha(k)=\hat \al(-k)$, $k\in\bbT$. In addition, $\hat\alpha\in
   C^{\infty}(\bbT)$ and it can be written as
 \begin{equation}
\label{om2bb}
\hat \al(k)=4\tau_1{\frak s}^2( k)\varphi^2({\frak s}^2( k)).
\end{equation}
The function
$\varphi:[0,+\infty)\to(0,+\infty)$ is  smooth  and
  $\varphi(0)=1$.  For the abbreviation sake, we shall use the notation
\begin{equation}
\label{021701}
\frak s(k):=\sin(\pi k)\quad \mbox{and}\quad  \frak c(k):=\cos(\pi
k),\quad k\in\bbT.
\end{equation}
 The parameter
$\tau_1$ appearing in \eqref{om2bb}, called the {\em sound speed}, is defined  by 
\begin{equation}
\label{tau}
\tau_1:=\frac{\hat \al''(0)}{8\pi^2}.
\end{equation}
Let also
\begin{equation}
\label{phi-is}
   \varphi_-:=\inf\varphi\quad \mbox{and}\quad \varphi_+:=\sup\varphi.
\end{equation}





Since the Hamiltonian is invariant under the translations of the positions
${\frak q}_x$ of the atoms, the latter are not well defined, and the
configuration space of our system is given by 
$((\frak r_x, \frak p_x))_{x\in\Z}$, where $\frak r_x$ should be
thought as the distance between particle $x$ and $x-1$. 
The evolution is given by a system of stochastic differential
equations 
\begin{eqnarray}
&d{\frak r}_{x}(t) &=\nabla^*{\frak p}_x(t)\; dt
\label{eq:bas2},\\
&&\nonumber\\
& d{\frak p}_x(t)&=\left\{-\alpha*{\frak
    q}_{x}(t)-\frac{\ga}{2}(\beta*{\frak p}(t))_x\right\}dt \nonumber\\
&&\quad +\ga^{1/2}\sum_{z=-1,0,1}(Y_{x+z}{\frak p}_x(t))dw_{x+z}(t),\quad x\in\bbZ.\nonumber
\end{eqnarray}
with the parameter $\ga>0$ that determines the strength of the noise
in the system. The   vector fields  $(Y_x)_{x\in\bbZ}$ are given by
\begin{equation}
\label{011210}
Y_x:=({\frak p}_x-{\frak p}_{x+1})\partial_{{\frak p}_{x-1}}+({\frak p}_{x+1}-{\frak p}_{x-1})\partial_{{\frak p}_{x}}+({\frak p}_{x-1}-{\frak p}_{x})\partial_{{\frak p}_{x+1}}.
\end{equation}
Here $(w_x(t),\,t\ge0)_{x\in\bbZ}$ are i.i.d. one dimensional, real
valued, standard Brownian motions,  that are  non-anticipative over
some filtered probability space 
$(\Om,{\cal F},\left({\cal F}_t\right),\bbP)$.  The symbol $*$ denotes
the convolution, but in particular, since
  $\hat\al(0)=0$, we can make sense of {(remember that ${\frak q}_x$ are not really defined)}
    $$
\alpha*{\frak
    q}_x:=
    \sum_{x'}
\alpha_{x-x'}{\frak
    q}_{x',x},
$$
where
\begin{equation}
\label{diff-q}
\frak
  {\frak
    q}_{x',x}:=\left\{
\begin{array}{ll}
\sum_{x'<x''\le x}\frak r_{x''},&\mbox{ if }x'<x,\\
&\\
0,&\mbox{ if }x'=x,\\
&\\
\sum_{x<x''\le x'}\frak r_{x''},&\mbox{ if }x<x'.
\end{array}
\right.
\end{equation}

Furthermore, 
 $\beta_x=\Delta\beta^{(0)}_x$ and $
\beta^{(1)}_x:=\nabla^*\beta^{(0)}_x,
$ where   
\begin{equation}
\label{022402}
 \beta^{(0)}_x=\left\{
 \begin{array}{rl}
 -4,&x=0,\\
 -1,&x=\pm 1,\\
 0, &\mbox{ if otherwise.}
 \end{array}
 \right.
 \end{equation}
The lattice Laplacian of a given $(g_x)_{x\in\bbZ}$ is defined as
  $\Delta g_x:=g_{x+1}+g_{x-1}-2g_x$. Let also $ \nabla g_x:=g_{x+1}-g_x$ and $ \nabla^*
g_x:=g_{x-1}-g_x$. 
A simple calculation shows that
\begin{equation}
\label{beta}
\hat \beta(k)=8\frak s^2( k)\left(1+2\frak c^2(k)\right)=8\frak s^2(
k)+4\frak s^2(2 k),
\end{equation}
and
\begin{equation}
\label{beta-1}
\hat \beta^{(1)}(k)=4ie^{-2i\pi k}{\frak s}(k)
\left(1+2\frak c^2(k)\right).
\end{equation}



The evolution equations are formulated on the Hilbert space $\ell_2$
made of all real valued sequences 
$\left(({\frak r}_x,{\frak p}_x)\right)_{x\in\bbZ}$ 
that satisfy
\begin{equation}
\label{ell-2}
 \sum_{x}\left({\frak r}_x^2+ {\frak
  p}_x^2\right)<+\infty.
\end{equation}

Concerning the initial data we assume that it  is distributed according to a 
probability measure
$\mu_\eps$  that
satisfies 
\begin{equation}
\label{011605}
\sup_{\eps\in(0,1]}\eps \sum_{x}\left\langle{\frak r}_x^2+ {\frak
  p}_x^2\right\rangle_{\mu_\eps}<+\infty.
\end{equation}
The above means that the total macroscopic  energy of the system is finite.
Here $\langle\cdot\rangle_{\mu_\eps}$ denotes the average with respect to $\mu_\eps$.

Denote by   $\bbE_\eps$  the expectation with respect to the
product measure
$\bbP_\eps=\mu_{\eps}\otimes \bbP$. 
The existence and uniqueness of a solution to \eqref{eq:bas2} in
$\ell_2$, with the aforementioned initial condition can be easily
concluded from the standard Hilbert space theory of stochastic differential
equations, see e.g. Chapter 6 of \cite{daza}.

\subsection{Energy density functional}
For a configuration  that satisfies \eqref{ell-2}, 
  the energy per
atom functional can be defined by:
\begin{equation}
  \label{eq:energy-a}
  \frak e_x ({\frak r}, {\frak p}):= \frac{\frak p_x^2}2 - \frac 14 \sum_{x'} \alpha_{x-x'} \frak
  q_{x,x'}^2.
\end{equation}
Notice that  $\sum_x \frak e_x  = \mathcal H({\frak q},{\frak p})$. We highlight here
the fact that, although the total energy is non-negative, in light of the assumptions a1)-a3) made about the interaction potential, the energy per atom $\frak e_x$ does not have definite sign. However we can prove, see Appendix \ref{appA} below, the following fact.
\begin{prop}
\label{prop011706}
We have
\begin{equation}
\label{equiv1}
c_-\sum_x ({\frak r}_x^2 +{\frak p}_x^2)\le 
 \sum_x {\frak e}_x ({\frak r}, {\frak p})
 \le c_+\sum_x ({\frak r}_x^2 +{\frak p}_x^2),\quad({\frak r}, {\frak p})\in\ell_2,
\end{equation}
with $c_-:=\min\{1/2,\tau_1\varphi_-^2\}$ and $c_+:=\max\{1/2,\tau_1\varphi_+^2\}$. In addition,
\begin{equation}
\label{equiv}
 c_*\sum_x |{\frak e}_x ({\frak r}, {\frak p})|\le 
 \sum_x {\frak e}_x ({\frak r}, {\frak p})
 \le \sum_x |{\frak e}_x ({\frak r}, {\frak p})|
\end{equation}
and
$
c_*:=c_-\left(\max\left\{1/2,\sum_{z>0}z^2|\al_z|\right\}\right)^{-1}.
$
\end{prop}

Thanks to \eqref{011605} and \eqref{equiv1} the distribution of the initial data is such that
the macroscopic energy of the chain at time $t=0$ is finite, i.e. that
\begin{equation}
\label{finite-energy0}
K_0:=\sup_{\eps\in(0,1]}\eps\sum_{x}\langle\frak e_x\rangle_{\mu_\eps}<+\infty.
\end{equation}
We shall also assume that there exists a macroscopic profile 
for the family of measures $\left(\mu_\eps\right)_{\eps>0}$, i.e. functions $ r(y)$, $ p(y)$ and $ e(y)$   belonging to $C_0^{\infty}(\bbR)$
such that the respective  $ w^T(y)=[ r(y),  p(y), e(y)]$ satisfies \eqref{010702x}.

\subsection{Macroscopic profiles of temperature and phononic energy}
The {\em macroscopic temperature profile} corresponding to the family of laws  $(\mu_\eps)_{\eps\in(0,1]}$ 
is defined as a function $T:\bbR\to[0,+\infty)$ such that
\begin{equation}
\label{T}
\int_{\bbR}T(y)J(y)dy=\lim_{\eps\to0+}\eps\sum_xJ(\eps x)\left\{\left\langle (\delta_\eps{\frak p}_x)^2\right\rangle_{\mu_\eps}
- \frac 14 \sum_{x'} \alpha_{x-x'} \left\langle(\delta_\eps\frak
  q_{x,x'})^2\right\rangle_{\mu_\eps}\right\}
\end{equation}
for any $J\in C_0^\infty(\bbR)$, where 
$\delta_\eps{\frak p}_x:={\frak p}_x- p(\eps x)$ and 
\begin{equation}
\label{diff-dq}
\delta_\eps\frak
  {\frak
    q}_{x',x}:=\left\{
\begin{array}{ll}
\sum_{x'<x''\le x}\delta_\eps\frak r_{x''},&\mbox{ if }x'<x,\\
&\\
0,&\mbox{ if }x'=x,\\
&\\
\sum_{x<x''\le x'}\delta_\eps\frak r_{x''},&\mbox{ if }x<x',
\end{array}
\right.
\end{equation}
with $\delta_\eps{\frak r}_x:={\frak r}_x- r(\eps x)$.
 After a simple calculation
    one gets the following identity
\begin{equation}
\label{e-bar}
T(y)= e(y)- e_{\rm ph}(y),
\end{equation}
with
\begin{equation}
\label{bar-e1}
 e_{\rm ph}(y):=\frac12\left( p^2(y)+\tau_1 r^2(y)\right) ,\quad
  y\in\bbR,
\end{equation}
the {\em phononic} (or {\em mechanical}) macroscopic energy profile.

\subsection{Energy spectrum}
The energy spectrum of the configuration distributed according to $\mu_\eps$  is defined as 
\begin{equation}
\label{022403}
{\frak w}_\eps(k):=\left\langle |\hat{\frak p}(k)|^2\right\rangle_{\mu_\eps}+\frac{\hat
    \al(k)}{4{\frak s}^2(k)} \left\langle |\hat{\frak
      r}(k)|^2\right\rangle_{\mu_\eps},
\end{equation}
where $\hat{\frak p}(k)$ and $\hat{\frak r}(k)$
are the Fourier transforms of $({\frak p}_x)$ and $({\frak
  r}_x)$, respectively.
Using \eqref{om2bb} we get
\begin{equation}
\label{022403a}
{\frak w}_\eps(k)
=\left\langle |\hat{\frak p}(k)|^2\right\rangle_{\mu_\eps}+\tau_1 \varphi^2({\frak s}^2( k))\left\langle |\hat{\frak
      r}(k)|^2\right\rangle_{\mu_\eps}
,\quad k\in\bbT.
\end{equation}

Assumption \eqref{finite-energy0} is equivalent with
\begin{equation}\label{finite-energy10}
 K_0=   \sup_{\eps\in(0,1]}\eps\int_{\bbT}{\frak w}_\eps(k)dk<+\infty.
\end{equation}
\begin{df}
\label{thermal}
{\em The  family of distributions $\left(\mu_\eps\right)_{\eps>0}$ is
  said to be of  {\em a thermal
type} if its energy spectrum ${\frak w}_\eps(k)$ satisfies 
\begin{equation}
\label{finite-energy1}
K_1:=\sup_{\eps\in(0,1]}\eps^2\int_{\bbT}{\frak  w}_{\eps}^2(k)dk<+\infty.
\end{equation}}
\end{df}
{\bf Remark.} In Section \ref{sec3.6}   we show that  \eqref{finite-energy1} implies that the respective macroscopic profiles $( r(y), p(y))$  vanish.   
Therefore, we conclude that  then the
macroscopic phononic energy $e_{\rm ph}(y)\equiv0$ (see
\eqref{bar-e1}) and, as a result,
\begin{equation}
\label{ini}
\lim_{\eps\to0+}\eps\sum_x J(\eps x) \langle {\frak
  e}_x\rangle_{\mu_{\eps}}=\int_{\bbR}J(y)T(y)dy,\quad J\in C_0^\infty(\bbR).
\end{equation}

We stress that although $ r(y)\equiv0$ and $ p(y)\equiv 0$, in
case condition \eqref{finite-energy1} holds, configuration $({\frak r},{\frak
  p})$ need not  necessarily be centered (in the measure $\mu_\eps$).
Condition \eqref{finite-energy1} is
related to the issue of variability of the initial data  on the 
microscopic scale, as can be seen in examples presented in Section \ref{sec10}.

\begin{df}
\label{phononic}
{\em The  family $\left(\mu_\eps\right)$ is called to be of  {\em a phononic (or mechanical)
type} if 
\begin{equation}
\label{010702}
\lim_{\eps\to0+}\eps\sum_x\left\langle\left[{\frak
    r}_x- r(\eps x)\right]^2\right\rangle_{\mu_{\eps}}=0
\end{equation}
and
\begin{equation}
\label{010702a}
\lim_{\eps\to0+}\eps\sum_x\left\langle\left[{\frak
    p}_x- p(\eps x)\right]^2\right\rangle_{\mu_{\eps}}=0,
\end{equation}
where $ r(x)$ and $ p(x)$ are the macroscopic profiles of the strain and momentum, see \eqref{010702x}.}
\end{df}

\bigskip
{\bf Remark.}
Using  \eqref{T} and \eqref{010702} together with \eqref{010702a} it
is straightforward to see  that the temperature profile corresponding
to a phononic type initial data vanishes, i.e.   $T(y)\equiv0$.

\bigskip

\section{Statement of the main results} 

\label{sec3a}

\bigskip

\subsection{Hyperbolic scaling}
\label{sec3a1}

 Assume that the configuration $({\frak r},{\frak  p})$ 
 can be decomposed into two parts whose laws  are respectively of  the thermal
 and phononic types. 
More precisely, we assume that
\begin{equation}
\label{012403}
{\frak r}_x={\frak r}_x'+{\frak r}_x'',\quad {\frak p}_x={\frak p}_x'+{\frak p}_x'',\quad x\in\bbZ,
\end{equation}
where
\begin{equation}
\label{finite-energy2}
K_1':=\sup_{\eps\in(0,1]}\eps^2\int_{\bbT}({\frak  w}_{\eps}')^2(k)dk<+\infty,
\end{equation}
and ${\frak w}_{\eps}'(k)$ is the energy spectrum of  $({\frak r}',{\frak p}')$, i.e.
\begin{equation}
\label{011611}
{\frak w}_{\eps}'(k)=\left\langle |\hat{\frak p}'(k)|^2\right\rangle_{\mu_\eps}+\tau_1 \varphi^2({\frak s}^2( k)) \left\langle |\hat{\frak r}'(k)|^2\right\rangle_{\mu_\eps}
,\quad k\in\bbT.
\end{equation}
with $(\hat{\frak p}'(k),\hat{\frak r}'(k))$
 the respective Fourier transforms of $({\frak p}',{\frak
  r}')$. In addition, we assume that the configuration $({\frak
  r}'',{\frak  p}'')$ is of the phononic type in the sense of
Definition \ref{phononic}.

  Suppose that $({\frak r}'(t),{\frak p}'(t))$ and  $({\frak r}''(t),{\frak p}''(t))$ describe the evolution of the respective initial configurations
 $({\frak r}'_x,{\frak
  p}'_x)$ and $({\frak r}_x'',{\frak
  p}_x'')$   under the dynamics \eqref{eq:bas2}.
Define the microscopic temperature and phononic energy density profiles by
\begin{equation}
\label{050306}
{\frak e}_{{\rm th},x}(t):={\frak e}_x({\frak r}'(t),{\frak p}'(t))
\quad\mbox{
and }\quad
{\frak e}_{{\rm ph},x}(t):={\frak e}_x({\frak r}''(t),{\frak p}''(t))
\end{equation}
Let
\begin{equation}
\label{040506}
 v^T(t,y):=( r(t,y), p(t,y))
\end{equation}
 be the solution of the linear wave equation
\begin{equation}
  \label{eq:wave:lin}
 \left\{
  \begin{array}{ll}
   & \partial_t  r(t,y) = \partial_y p(t,y),\\
    &\partial_t  p(t,y) = \tau_1\partial_y r(t,y), \\
    & r(0,y)=  r(y), \quad  p(0,y)= p(y).
      \end{array}
      \right.
\end{equation}
The macroscopic phononic energy at time $t$, see \eqref{bar-e1}, is given
by
\begin{equation}
\label{bar-e}
 e_{\rm ph}(t,y)=\frac12\left( p^2(t,y)+\tau_1 r^2(t,y)\right).
\end{equation}
The components of 
$$
 w^T(t,y) := \left[  r(t,y),  p(t,y),  e_{\rm
    ph}(t,y)\right]
$$
evolve according to the system of linear
Euler equations:
\begin{equation}
  \label{eq:euler:l}
 \left\{
  \begin{array}{ll}
    &\partial_t  r(t,y) = \partial_y p(t,y),\\
    &\partial_t  p(t,y) = \tau_1\partial_y r(t,y), \\
    &\partial_t  e_{\rm ph}(t,y) = \tau_1\partial_y\left( r(t,y)
      p(t,y)\right),\\
      & w(0,y)= w(y).
  \end{array}
  \right.
\end{equation}
The following result is proven in Section \ref{sec-phon1}.
\begin{thm}
\label{cor011002}
Suppose that the initial configuration $ ({\frak r},{\frak
  p})$ satisfies the assumptions formulated in the foregoing. Then,
$({\frak r}'(t\eps^{-1}),{\frak p}'(t\eps^{-1}))$ is of  thermal type for all $t\ge0$ and
\begin{equation}
\label{011002-a}
\lim_{\eps\to0+}\eps\sum_x\int_0^{+\infty}J(t,\eps x) \bbE_\eps {\frak e}_{{\rm th},x}\left(\frac{t}{\eps}\right)dt=\int_0^{+\infty}dt\int_{\bbR}T(y)J(t,y)dy 
\end{equation}
 for any $J\in
C_0^{\infty}([0,+\infty)\times \bbR)$.
Configurations
 $({\frak r}''(t\eps^{-1}),{\frak p}''(t\eps^{-1}))$ are of  phononic type for all $t\ge0$ and
\begin{equation}
\label{011002-b}
\lim_{\eps\to0+}\eps\sum_x\left|\bbE_\eps {\frak e}_{{\rm
    ph},x}\left(\frac{t}{\eps}\right)- e_{\rm
  ph}(t,\eps x)\right|=0.
\end{equation}
 In addition,
\begin{equation}
\label{011002-d}
\lim_{\eps\to0+}\eps\sum_x\int_0^{+\infty}J(t,\eps x) \bbE_\eps {\frak
  e}_{x}\left(\frac{t}{\eps}\right)dt=\int_0^{+\infty}\int_{\bbR}\left[ e_{\rm ph}(t,y)+T(y)\right]J(t,y)  dtdy,
\end{equation}
for any $J\in
C_0^{\infty}([0,+\infty)\times \bbR)$.
\end{thm}

\bigskip
Concerning the macroscopic evolution of  the vector  $
\frak w_x (t)$, see \eqref{w}, the above result implies  the following.
 \begin{cor}
\label{cor031102}
 Assume  that 
\begin{equation}
\label{020301-15a}
  \lim_{\eps\to0+} 
 \eps \sum_x J(\eps x) \langle\frak w_x\left(0\right)\rangle_{\mu_\eps}
  =
   \int_{\bbR} J(y)  w(y) \; dy,\quad J\in C_0^\infty(\bbR).
\end{equation}
Then, for any  $J\in C_0^\infty([0,+\infty)\times \bbR)$ we have
\begin{equation}
\label{020301-15}
  \lim_{\eps\to0+} 
\eps \sum_x\int_0^{+\infty} J(t,\eps x)  \bbE_\eps\frak w_x\left(\frac{ t}{\eps}\right)dt
  =
  \int_0^{+\infty} \int_{\bbR} J(t,y)  w(t,y)dt dy,\qquad 
\end{equation}
with the components of $ w(t,y)$ being the solutions of
the linear Euler system \eqref{eq:euler:l}.
\end{cor}

\bigskip

\subsection{Super-diffusive scaling}

It follows from Theorem \ref{cor011002} that  the phononic energy evolution takes place on 
the hyperbolic scale that is described by the linear Euler equation. In
consequence, it gets dispersed to infinity on  time scales longer than
$t\eps^{-1}$.
On the other hand, it has  been shown, see 
Theorem 3.1 of   \cite{JKO}, that the respective temperature profile
$T(y)$, see \eqref{T}, 
evolves on a superdiffusive scale $(t\eps^{-3/2},x\eps^{-1})$. 
Therefore, we have  the following result.
\begin{thm}
\label{energy-prop-main}
 Suppose that the distribution of the initial configuration $ ({\frak r},{\frak
  p})$ satisfies 
condition \eqref{012403}.
 Then, for any test function $J\in 
C^{\infty}_0([0,+\infty)\times \bbR)$ we  have: 
\begin{equation}
\label{041803}
 \lim_{\eps\to0+}\eps\sum_{x}\int_0^{+\infty} J(t,\eps x)  \bbE_{\eps}{\frak
   e}_{x}\left(\frac{t}{\eps^{3/2}}\right) dt
=   \int_0^{+\infty}\int_{\bbR}T(t,y) J(t,y)dtdy.
 \end{equation}
Here $T(t,y)$ satisfies the fractional heat equation \eqref{eq:frheat}
with the initial condition $T(0,y)=T(y)$, given by \eqref{ini}.
\end{thm}
The  proof of the above result is presented in  Section \ref{sec-phon}.

\subsection{Evolution of the normal modes at the diffusive time scale}

Finally we 
consider 
the Riemann invariants of the linear wave equation system
\eqref{eq:wave:lin}. They are given by
\begin{equation}
\label{021901-15}
f^{(\pm)}(t,y):=\bar
p(t,y)\pm\sqrt{\tau_1}\bar
r(t,y).
\end{equation}
The quantities defined in \eqref{021901-15} are constant along the
characteristics of   \eqref{eq:wave:lin}, which  are  given by the straight lines $y\pm \sqrt{\tau_1} t={\rm const}$. Therefore 
\begin{equation}
\label{041703}
f^{(\pm)}(t,y)=f^{(\pm)}(y\pm \sqrt{\tau_1} t),
\end{equation}
with $f^{(\pm)}(y)$ determined from the initial datum.
This motivates the introduction of the microscopic {\em normal modes}
given by
\begin{eqnarray}
\label{phonon-1}
&&
\frak f^{(+)}_y(t):= {\frak p}_y(t)+\sqrt{\tau_1}
\left[1+\frac12\left(\frac{3\ga}{\sqrt{\tau_1}}-1\right)\nabla^*\right]{\frak
  r}_y(t), \nonumber\\
&&\\
&&
\frak f^{(-)}_y(t):= {\frak p}_y(t)-\sqrt{\tau_1}
\left[1-\frac12\left(\frac{3\ga}{\sqrt{\tau_1}}+1\right)\nabla^*\right]{\frak
  r}_y(t),\nonumber
\end{eqnarray}
that are the second order approximations (up to a diffusive scale) of
$f^{(\pm)}(t,y)$. 
The particular form of $\frak
f^{(\pm)}(t)$ is determined by the fact that we are looking for
quantities  
of the form $ {\frak p}(t)\pm \sqrt{\tau_1}
\left(1\pm c_{\pm}\nabla^*\right){\frak
  r}(t)$ 
that, at the hyperbolic scale,  approximate the Riemann invariants \eqref{021901-15}
and  possess limits at the diffusive time scale. The
latter requirement determines  that $c_\pm =(1/2)\left(3\ga/\sqrt{\tau_1}\pm1\right)$.

\bigskip

{\bf Remark.} The  normal modes $\frak f^{(\pm)}(t)$ capture the
fluctuations around the Riemann invariants of the linear hyperbolic
system \eqref{eq:euler}. Their analogues, called normal modes
$\phi_{\pm}(t)$, are  considered in Appendix 1 b) of \cite{Sp13}, in
the context of anharmonic chains with FPU-potential, see
(8.18) of ibid. 

\bigskip

 To state our  result rigorously,  assume that
 $\left(\mu_\eps\right)_\eps$ - the initial laws  of configurations
 $({\frak r},{\frak p})$  satisfy \eqref{010702x}.
Denote also the heat kernel
\begin{equation}
\label{p-t}
P_t(y):=\frac{1}{\sqrt{4\pi Dt}}\exp\left\{-\frac{y^2}{4Dt}\right\},
\end{equation}
where  $
D:=3\ga
$. Let 
$
f^{(d)}_{\iota}(t,y):=P_t*f^{(\iota)}(y)
$, $\iota=\pm$  
and
\begin{equation}
\label{012504}
f^{(\pm)}(y):=p(y)\pm\sqrt{\tau_1}r(y).
\end{equation}
 With the above notation we can formulate the following result. 
\begin{thm}
\label{p-modes}
Suppose that condition \eqref{010702x} is in force. 
Then, the phonon modes $f^{(\pm)}_y(t)$ satisfy
\begin{equation}
\label{u1}
\lim_{\eps\to0+}\eps\sum_x J(\eps x)\bbE_{\eps}{\frak
  f}_x^{(\pm)}\left(\frac{t}{\eps}\right)=\int_{\bbR}J(y)f^{(\pm)}(t,y)dy.
\end{equation}
for any $J\in C_0^\infty(\bbR)$. In addition, for any $\iota\in\{-,+\}$
\begin{equation}
\label{ud1}
\lim_{\eps\to0+}\eps\sum_{x}J\left(\eps x-\iota\sqrt{\tau_1}\frac{t}{\eps}\right) \bbE_{\eps}{\frak
  f}^{(\iota)}_x\left(\frac{t}{\eps^2}\right)=\int_{\bbR}f^{(d)}_{\iota}(t,y)J(y)dy,
\end{equation}
\end{thm}
The proof of this theorem is contained in  Section \ref{sec9}.

\bigskip

{{\bf Remark.} {The results of the present paper are valid also
for other  stochastic dynamics obtained by a stochastic perturbation of the harmonic chain that conserves
the volume,  energy and momentum.
For example we can take  a  harmonic chain perturbed by a noise of the ''jump'' type.}
More precisely, let $( N_{x,x+1}(t))_{x\in\bbZ} ~~$ be
i.i.d.  Poisson processes with  intensity $3\ga/2$.
The dynamics of the strain component $\left({\frak r}_x(t)\right)_{x\in\bbZ}$  is the same
as in \eqref{eq:bas2}, while the momentum 
$\left({\frak p}_x(t)\right)_{x\in\bbZ}$ is a c\`adl\`ag process given by
\begin{eqnarray}
\label{eq:bas1a}
d{\frak p}_x(t)=&& -(\alpha*{\frak q}(t))_xdt\nonumber\\
&&
\\
&&+\left[\nabla{\frak p}_x(t-)d N_{x,x+1}(t)+\nabla^*{\frak p}_x(t-)dN_{x-1,x}(t)\right],\quad x\in\bbZ\nonumber
\end{eqnarray}
where $(\alpha*{\frak q})_x(t)$ is defined as in \eqref{diff-q}. One can show, with  essentially the same arguments as the ones used in the
present paper, that Theorems  \ref{cor011002} -- \ref{p-modes} hold also for this dynamics.}

\bigskip

\section{Asymptotics of the phononic ensemble}

\label{sec4.0}

In this section we assume that the laws $(\mu_\eps)_{\eps>0}$ of
the initial configuration $({\frak r},{\frak p})$  is of the
phononic type, i.e. they satisfy \eqref{010702} and \eqref{010702a}.

\subsection{Approximation of the macroscopic phononic energy}
 The solution of \eqref{eq:wave:lin} is given by
\begin{eqnarray}
\label{063003}
&&
 r(t,y) = r\left(y+\sqrt{\tau_1} t\right)+
r\left(y-\sqrt{\tau_1} t\right)+\frac{1}{\sqrt{\tau_1}}\left[
  p\left(y+\sqrt{\tau_1} t\right)- p\left(y-\sqrt{\tau_1}
    t\right)\right]\nonumber\\
&&
\\
&&
 p(t,y) = p\left(y+\sqrt{\tau_1} t\right)+
p\left(y-\sqrt{\tau_1} t\right)+\sqrt{\tau_1}\left[
  r\left(y+\sqrt{\tau_1} t\right)- r\left(y-\sqrt{\tau_1}
    t\right)\right],\nonumber
\end{eqnarray}
where  $ r(y), p(y)$ are the initial data in
\eqref{eq:wave:lin}. 
Using the above formulas and \eqref{bar-e} we can write  $
  e_{\rm ph}\left(t,y\right)$ in terms of the profiles $ r(y)$ and
  $ p(y)$.
We can easily infer the following.
\begin{lemma}
\label{lm010606}
Suppose 
that $ r, p\in C_0^\infty(\bbR)$. Then,
\begin{equation}
\label{080506}
\lim_{\eps\to0+}\sup_{t\ge0}\left|\eps
\sum_x
  e_{\rm ph}\left(t,\eps x\right)-\int_{\bbR} e_{\rm ph}(t,y)dy\right|=0.
\end{equation}
\end{lemma}

\bigskip

\subsection{Evolution of the mean ensemble}

Define the  mean configuration
$$
(\bar{\frak r}_\eps(t), \bar{\frak
  p}_\eps(t))= (\bar{\frak r}_{\eps,x}(t), \bar{\frak
  p}_{\eps,x}(t))_{x\in\bbZ},
$$
 where  
  \begin{equation}
\label{mean}
  \bar{\frak r}_{\eps,x}(t):=\bbE_\eps {\frak r}_{x}(t),\quad    \bar{\frak p}_{\eps,x}(t):=\bbE_\eps {\frak p}_{x}(t).
 \end{equation}
The  respective energy  density 
is defined then as 
\begin{equation}
  \label{eq:phon-energy}
  \bar{\frak e}_{x}^{(\eps)}(t) := {\frak e}_x(\bar{\frak r}_\eps(t), \bar{\frak
  p}_\eps(t))
 = \frac{\bar{\frak p}^2_{\eps,x}(t)}2 - \frac 14 \sum_{x'} \alpha_{x-x'} \bar{\frak q}_{\eps,x,x'}^2 (t),
\end{equation}
where  $\bar{\frak q}_{\eps,x,x'} (t)$ is the  respective mean of ${\frak q}_{x,x'} (t)$ defined by \eqref{diff-q}.

Let
  $\hat{\frak v}^T(t,k):=[\hat{\frak r}(t,k),\hat{\frak
  p}(t,k)]$, $k\in\bbT$,  where $\hat{\frak r}(t,k),\hat{\frak
  p}(t,k)$  are the Fourier transforms of the respective
   components of the configuration $({\frak r}_x(t),{\frak p}_x(t)))$.
Suppose that $\delta\ge 1$. Define 
\begin{equation}
\label{R-eps}
\hat{\bar{\frak r}}_\eps(t,k):=\eps \bbE_\eps\hat {\frak
  r}\left(\frac{t}{\eps^{\delta}},\eps k\right)\quad
\mbox{and}\quad \hat{\bar{\frak p}}_\eps(t,k):=\eps \bbE_\eps\hat{\frak p}\left(\frac{t}{\eps^{\delta}},\eps k\right).
\end{equation}
Conservation of energy and condition \eqref{finite-energy0} imply that
\begin{equation}
\label{052402}
E_*:=\sup_{t\ge0, \eps\in(0,1]}\int_{\eps^{-1}\bbT}(|\hat{\bar{\frak r}}_\eps(t,k)|^2+|\hat{\bar{\frak p}}_\eps(t,k)|^2)dk<+\infty.
\end{equation}
From \eqref{eq:bas2} we obtain
\begin{equation}
\label{A-eps}
\frac{d}{dt}\hat{\bar{\frak v}}_\eps(t,k)=\frac{1}{\eps^{\delta-1}}A_\eps(k) \hat{\bar {\frak v}}_\eps(t,k),
\end{equation}
where 
\begin{equation}
\label{V-eps}
\hat{\bar{\frak v}}_\eps^T(t,k)=[\hat{\bar{\frak r}}_\eps(t,k),\hat{\bar{\frak
  p}}_\eps(t,k)]
\end{equation}
 and 
\begin{equation}
\label{A-eps1}
A_\eps(k):=\left[
\begin{array}{ccc}
0 & &a\\
&&\\
-\tilde\tau(\eps k)a^*&&-b
\end{array}\right],
\end{equation}
with $a^*$ - the complex conjugate of $a$ and
\begin{eqnarray}
\label{010212}
&&
\tilde \tau( k):=\tau_1\varphi^2(\sin^2(\pi  k)),\quad a:=\frac{1}{\eps}(1-\exp\left\{-2\pi i\eps k\right\}),
\nonumber\\
&&
b:=\frac{\ga}{\eps}|1-\exp\left\{-2\pi i\eps k\right\}|^2[2+\cos(2\pi \eps k)].
\end{eqnarray}
Observe that for a given $M>0$ we have
\begin{equation}
\label{022502}
\lim_{\eps\to0+}\sup_{|k|\le M}|A_\eps(k)-A_0(k)|=0
\end{equation}
and
$$
A_0(k):=\left[
\begin{array}{ccc}
0 & &2\pi i k\\
&&\\
2\pi i\tau_1 k&&0
\end{array}\right].
$$
Since both eigenvalues of $A_0(k)$ are imaginary we have
\begin{equation}
\label{042502}
D_*:=\sup_{(t,k)\in\bbR^2}\|\exp\left\{A_0(k)t\right\}\|<+\infty.
\end{equation}
Here $\|\cdot\|$ is the matrix norm defined as the norm of the
corresponding linear operator on a euclidean space.
In fact,  an analogue of \eqref{042502} holds in the case of the linear dynamics governed by \eqref{A-eps}.
\begin{lm}
\label{lm012402}
We have
\begin{equation}
\label{012402}
C_*:=\sup_{(t,k)\in\bbR^2,\eps\in(0,1]}\|\exp\left\{A_\eps(k)t\right\}\|<+\infty.
\end{equation}
\end{lm}
\proof The eigenvalues of $A_\eps(k)$ equal
$$
\la_{\pm}=\frac{-1}{2}\left\{b\pm\sqrt{b^2-4\tilde\tau(\eps k)|a|^2}\right\}.
$$
Since both $b$ and $\tilde\tau(\eps k)$ (see \eqref{010212}) are real and non-negative it is clear that ${\rm Re}\la_{\pm}\le 0$ and the conclusion of the lemma follows.
\qed

\subsection{Asymptotics of  the mean ensemble}

\label{sec4.1}

Let $\hat{\bar{\frak v}}_\eps^{(0)}(t,k)$ be the solution of the
following equation
\begin{equation}
\label{072302}
\frac{d}{dt}\hat{\bar{\frak v}}_\eps^{(0)}(t,k)=\frac{1}{\eps^{\delta-1}}A_0(k) \hat{\bar {\frak v}}_\eps^{(0)}(t,k),
\quad \hat{\bar{\frak v}}_\eps^{(0)}(0,k) =\hat{\bar{\frak v}}_\eps(0,k).
\end{equation}
Let also $\hat v(t,\ell)$ be the Fourier transform of $v(t,y)$ the solution of
the linear wave equation system given in \eqref{040506}.
It satisfies
\begin{equation}
\label{050506}
\hat v(t,\ell)=\exp\left\{A_0(\ell)t\right\}\hat v(0,\ell).
\end{equation}
Conditions \eqref{010702} and \eqref{010702a} imply
\begin{equation}
\label{032402bb}
\lim_{\eps\to0+}\int_{\eps^{-1}\bbT}\left|\hat{\bar{\frak
  v}}_{\eps}(0,\ell)-\hat v(0,\ell)\right|^2d\ell=0.
\end{equation}
\begin{lm}
\label{lm012602}
Suppose that $\delta\in[1,2)$ and $M>0$. Then,
\begin{equation}
\label{012502a}
\lim_{\eps\to0+}\int_{-M}^M|\hat{\bar{\frak v}}_\eps(t,k) -\hat{\bar{\frak v}}_\eps^{(0)}(t,k)|^2dk=0,\quad t\ge0.
\end{equation}
\end{lm}
\proof Note that
$
A_\eps(k)=A_0(k)+\eps B_\eps (k)
$ (see \eqref{A-eps1}), where 
\begin{equation}
\label{032602}
b_*(M):=\sup_{\eps\in(0,1]}\sup_{|k|\le M}\|B_\eps (k)\|<+\infty.
\end{equation}
From \eqref{A-eps} we can write
\begin{equation}
\label{A-eps2}
\frac{d}{dt}\hat{\bar{\frak v}}_\eps(t,k)=\frac{1}{\eps^{\delta-1}}A_0(k) \hat{\bar {\frak v}}_\eps(t,k)+\eps^{2-\delta}B_\eps(k) \hat{\bar {\frak v}}_\eps(t,k),
\end{equation}
Therefore, by Duhamel's formula, we conclude
\begin{equation}
\label{duhamel1}
\hat{\bar{\frak v}}_\eps(t,k) -\hat{\bar{\frak v}}_\eps^{(0)}(t,k)=\eps^{2-\delta}\int_0^t \exp\left\{\frac{t-s}{\eps^{\delta-1}}A_0(k)\right\} B_\eps (k)\hat{\bar{\frak v}}_\eps(s,k)dk.
\end{equation}
The  lemma then follows from \eqref{052402}, \eqref{032602} and
Lemma \ref{lm012402}.
\qed

\bigskip

\begin{lm}
\label{lm022602}
For any $t\ge0$ and $\delta\in[1,2)$ we have
\begin{equation}
\label{032402}
\lim_{\eps\to0+}\int_{\eps^{-1}\bbT}\left|\hat{\bar{\frak
  v}}_{\eps}(
t,\ell)-\hat v\left(\frac{t}{\eps^{\delta-1}},\ell\right)\right|^2d\ell=0
\end{equation}
and, in consequence,
\begin{equation}
\label{032402a}
\lim_{\eps\to0+}\eps\sum_x\left|\bar{\frak
  v}_{x}\left(\frac{t}{\eps^{\delta}}\right)-v\left(\frac{t}{\eps^{\delta-1}},\eps x\right)\right|^2=0.
\end{equation}
\end{lm}
\proof 
Formula \eqref{032402a} follows from \eqref{032402} so we only focus on the proof of the latter.
We have
$$
\hat{\bar{\frak
  v}}_{\eps}(t,\ell)=\exp\left\{A_\eps(\ell)\frac{t}{\eps^{\delta-1}}\right\}\hat{\bar{\frak
  v}}_{\eps}(0,\ell).
$$
Therefore,from the above and \eqref{050506}  we can write 
\begin{eqnarray}
\label{032502}
&&
\limsup_{\eps\to0+}\int_{\eps^{-1}\bbT}\left|\hat{\bar{\frak
  v}}_{\eps}(
t,\ell)-\hat v\left(\frac{t}{\eps^{\delta-1}},\ell\right)\right|^2d\ell \nonumber\\
&&
\le
2\limsup_{\eps\to0+}\int_{\eps^{-1}\bbT}\left|\exp\left\{A_\eps(\ell)\frac{t}{\eps^{\delta-1}}\right\}\left[\hat{\bar{\frak
  v}}_{\eps}(0,\ell)-\hat
v(0,\ell)\right]\right|^2d\ell
\\
&&
+2\limsup_{\eps\to0+}\int_{\eps^{-1}\bbT}\left|\exp\left\{A_\eps(\ell) \frac{t}{\eps^{\delta-1}}\right\}\hat
v(0,\ell)-\exp\left\{A_0(\ell) \frac{t}{\eps^{\delta-1}}\right\}\hat
v(0,\ell)\right|^2d\ell.\nonumber
\end{eqnarray}
Using Lemma \ref{lm012402} and \eqref{032402bb} we conclude that the
first term on the right hand side vanishes. To estimate the second
term divide the domain of integration into
the regions $|\ell|\le M$ and its complement. On the first region we
use Lemma \ref{lm012602} to conclude that the respective limit
vanishes. Therefore, we can write that, modulo multiplication by factor $2$, the second term  on the right
hand side of \eqref{032502} is  equal to
\begin{eqnarray}
\label{052502}
&&
\limsup_{\eps\to0+}\int_{M\le |\ell|\le 1/(2\eps)}\left|\exp\left\{A_\eps(\ell)\frac{t}{\eps^{\delta-1}}\right\}\hat
v(0,\ell)-\exp\left\{A_0(\ell)\frac{t}{\eps^{\delta-1}}\right\}\hat
v(0,\ell)\right|^2d\ell\nonumber\\
&&
\le 2 \limsup_{\eps\to0+}\int_{M\le |\ell|\le 1/(2\eps)}\left\{\left|\exp\left\{A_\eps(\ell)\frac{t}{\eps^{\delta-1}}\right\}\hat
v(0,\ell)\right|^2+\left|\exp\left\{A_0(\ell)\frac{t}{\eps^{\delta-1}}\right\}\hat
v(0,\ell)\right|^2\right\}d\ell \nonumber\\
&&
\le 4 \left(C_*+D_*\right)\limsup_{\eps\to0+}\int_{M\le |\ell|\le 1/(2\eps)}\left|\hat
v(0,\ell)\right|^2d\ell,
\end{eqnarray}
the last estimate following from Lemma \ref{lm012402} and estimate
\eqref{042502}. We can adjust $M>0$ to become sufficiently large so that the
utmost right hand side of \eqref{052502} can be arbitrarily small.  The conclusion of the
lemma therefore follows.
\qed

\bigskip

\bigskip

Suppose also that  $(r(t,y),p(t,y))$ is the solution of the
linear wave equation \eqref{eq:wave:lin} and
 $\bar
e_{\rm ph}(t,y)$ is  the corresponding macroscopic phononic energy density profile, defined  by \eqref{bar-e}. As a direct corollary from Lemma \ref{lm022602} we conclude the
following.
\begin{cor}
\label{thm1a}
For any $t\ge0$ we have
\begin{eqnarray}
\label{020702x}
&&
\lim_{\eps\to0+}\eps\sum_x \left[\bar {\frak r}_{\eps,x}\left(\frac{t}{\eps}\right)-r(t,\eps x)\right]^2=0,\nonumber
\\
&&
\\
&&
\lim_{\eps\to0+}\eps\sum_x \left[\bar {\frak p}_{\eps,x}\left(\frac{t}{\eps}\right)-p(t,\eps x)\right]^2=0.\nonumber
\end{eqnarray}
\end{cor}

Concerning the behavior of energy functional $\bar{\frak
  e}_{x}^{(\eps)}\left(t\right)$  corresponding to the
average phononic ensemble (see 
\eqref{eq:phon-energy}) we get the following.
\begin{cor}
\label{cor010506}
For any $t\ge0$ and $\delta\in[1,2)$ we have
\begin{equation}
\label{020506}
\lim_{\eps\to0+}\eps\sum_x\left|\bar{\frak
  e}_{x}^{(\eps)}\left(\frac{t}{\eps^{\delta}}\right)-e_{\rm ph}\left(\frac{t}{\eps^{\delta-1}},\eps x\right)\right|=0,
\end{equation}
where $e_{\rm ph}(t,y)$ is given by \eqref{bar-e}.
In addition, for any 
$\delta\in(1,2)$  we have
\begin{equation}
\label{010402-15}
\lim_{\eps\to0+}\eps\sum_x J(\eps x) \bar{\frak
  e}_{x}^{(\eps)}\left(\frac{t}{\eps^{\delta}}\right)=0,\quad J\in C_0^\infty(\bbR).
\end{equation}
\end{cor}
\proof
In light of 
Lemma \ref{lm022602}
only equality  (\ref{010402-15})
requires a proof.
Assume that $\delta\in(1,2)$. Note that
the left hand side of \eqref{010402-15}  equals
$\lim_{\eps\to0+}[{\cal J}_1^{(\eps)}(t)+{\cal J}_2^{(\eps)}(t)]$, where
{
\begin{eqnarray}
\label{012602}
&&
{\cal J}_1^{(\eps)}(t):=\frac{\eps}{2}\sum_x\bar{\frak p}^2_{\eps,x}\left(\frac{t}{\eps^{\delta}}\right)J(\eps
x)\\
&&
=\frac12\int_{-1/(2\eps)}^{1/(2\eps)}\tilde J_\eps\left(\ell\right)d\ell\int_{-1/(2\eps)}^{1/(2\eps)}\hat{\bar{\frak p}}_{\eps}\left(\frac{t}{\eps^{\delta-1}},-\ell-k\right)\hat{\bar{\frak p}}_{\eps}\left(\frac{t}{\eps^{\delta-1}},k\right) dk\nonumber
\end{eqnarray}}
and
{
\begin{eqnarray}
\label{022602}
&&
{\cal J}_2^{(\eps)}(t):= -\frac {\eps}{4}\sum_{x,x'}J(\eps x)
\alpha_{x-x'} \bar{\frak q}^2_{\eps,x}\left
    (\frac{t}{\eps^{\delta}}\right) =
-\sum_{z} \frac {\alpha_{z}}{4}\int_{-1/(2\eps)}^{1/(2\eps)}\tilde
J_\eps\left(\ell\right)d\ell\nonumber\\
&&
\\
&&
\times\int_{-1/(2\eps)}^{1/(2\eps)}h_z(-\eps(k+\ell))
h_z(\eps k)\hat{\bar{\frak r}}_{\eps}\left(\frac{t}{\eps^{\delta-1}},-\ell-k\right) \hat{\bar{\frak r}}_{\eps}\left(\frac{t}{\eps^{\delta-1}},k\right) dk,\nonumber
\end{eqnarray}
where
$$
h_z(k):=\frac{\exp\left\{-2\pi i k z\right\}-1}{\exp\left\{-2\pi i
    k\right\}-1},\quad z\in\bbZ,\,k\in\bbT
$$
and
$$
\tilde J_\eps\left(\ell\right):=\sum_m\hat
J\left(\ell+\frac{m}{\eps}\right).
$$}
Recall that $\hat{\bar{\frak v}}_\eps(t,k)$ and $\hat{\bar{\frak v}}_\eps^{(0)}(t,k)$ are
given by \eqref{A-eps} and \eqref{072302} respectively. 
Using   Lemma \ref{lm022602}, 
we conclude  that
$$
\lim_{\eps\to0+}[{\cal J}_1^{(\eps)}(t)-\bar{\cal J}_1^{(\eps)}(t)]=0\quad \mbox{and} \quad \lim_{\eps\to0+}[ {\cal J}_2^{(\eps)}(t)-\bar{\cal J}_2^{(\eps)}(t)]=0,
$$
where
\begin{equation}
\label{012602b}
\bar{\cal J}_1^{(\eps)}(t):=\int_{\bbR}\hat
J(\ell)d\ell\int_{\bbR}\hat
p\left(\frac{t}{\eps^{\delta-1}},-\ell-k\right)\hat p\left(\frac{t}{\eps^{\delta-1}},k\right) dk
\end{equation}
and
\begin{equation}
\label{022602b}
{\cal J}_2^{(\eps)}(t):= 
-\sum_z \frac {z^2\alpha_{z}}{4}\int_{\bbR}\hat
J(\ell)\int_{\bbR}\hat r\left(\frac{t}{\eps^{\delta-1}},-\ell-k\right) \hat
r\left(\frac{t}{\eps^{\delta-1}},k\right)d\ell dk.
\end{equation}
Recall that
$$
\hat p\left(\frac{t}{\eps^{\delta-1}},k\right)=\left\langle \exp\left\{A_0(k)\frac{t}{\eps^{\delta-1}}\right\}\hat
  v(k), {\rm
    e}_2\right\rangle_{\mathbb C^d}
$$
where ${\rm e}_2^T=[0,1]$ and $\langle\cdot,\cdot\rangle_{\mathbb C^d}$ is the scalar product on $\mathbb C^d$. Therefore
\begin{eqnarray*}
&&
\bar{\cal J}_1^{(\eps)}(t)=\int_{\bbR}\hat
J(\ell)d\ell\int_{\bbR}
\left\langle \hat
  v(-k-l),\exp\left\{A_0^T(-k-\ell)\frac{t}{\eps^{\delta-1}}\right\} {\rm
    e}_2\right\rangle_{\mathbb C^d}\\
&&
\left\langle \hat
  v(k),\exp\left\{A_0^T(k)\frac{t}{\eps^{\delta-1}}\right\} {\rm
    e}_2\right\rangle_{\mathbb C^d} dk.
\end{eqnarray*}
By an elementary application of the Riemann-Lebesgue theorem we get
$$
\lim_{\eps\to0+}\bar{\cal J}_j^{(\eps)}(t)=0,\quad j=1,2.
$$
This ends the proof of  \eqref{010402-15}.
\qed

\bigskip


\subsection{Evolution of the conserved quantities at the hyperbolic scale}
Our  goal in this section is to prove the  following
result.

\begin{thm}
\label{thm1}
Suppose that \eqref{010702} and \eqref{010702a} are in force.
Then,
\begin{eqnarray}
\label{020702}
&&
\lim_{\eps\to0+}\eps\sum_x\bbE_\eps\left[{\frak r}_x\left(\frac{t}{\eps}\right)-r(t,\eps x)\right]^2=0,\nonumber\\
&&
\\
&&
\lim_{\eps\to0+}\eps\sum_x \bbE_\eps\left[{\frak p}_x\left(\frac{t}{\eps}\right)-p(t,\eps x)\right]^2=0\nonumber
\end{eqnarray}
and 
\begin{equation}
\label{011002}
\lim_{\eps\to0+}\eps\sum_x\left| \bbE_{\eps}{\frak
    e}_{x}\left(\frac{t}{\eps}\right)-e_{\rm ph}(t,\eps
  x)\right|=0,\quad \,t\ge0.
\end{equation}
\end{thm}
\proof Equality \eqref{011002} is a consequence of \eqref{020702}.
We only need to substantiate the latter.
Let
$
(\delta{\frak r}_\eps(t), \delta{\frak
  p}_\eps(t)):= (\delta{\frak r}_{\eps,x}(t), \delta{\frak
  p}_{\eps,x}(t))_{x\in\bbZ},
$
  where
  \begin{equation}
\label{mean1}
   \delta{\frak r}_{\eps,x}(t):  ={\frak r}_{\eps,x}(t)- \bar{\frak r}_{\eps,x}(t),\quad    \delta{\frak p}_{\eps,x}(t):  ={\frak p}_{\eps,x}(t)- \bar{\frak p}_{\eps,x}(t),
 \end{equation}
with $\bar{\frak r}_{\eps,x}(t)$ and $\bar{\frak p}_{\eps,x}(t)$ given
by \eqref{mean}. We have (see \eqref{eq:phon-energy})
\begin{equation}
\label{052603}
\bbE_\eps{\frak e}_x({\frak r}_\eps(t), {\frak
  p}_\eps(t))=
\bbE_\eps{\frak e}_x(\delta{\frak r}_\eps(t), \delta{\frak
  p}_\eps(t))+
\bar{\frak e}_x^{(\eps)}\left(t\right) .
\end{equation}
By  conservation
of the total energy
we obtain
\begin{eqnarray}
\label{070506}
&&
\eps\sum_x\left\langle{\frak e}_x({\frak r}, {\frak
  p})\right\rangle_{\mu_\eps}=\eps\sum_x\bbE_\eps{\frak e}_x\left({\frak r}\left(\frac{t}{\eps}\right), {\frak
  p}\left(\frac{t}{\eps}\right)\right)\nonumber\\
&&
=
\eps\sum_x\bbE_\eps{\frak e}_x\left(\delta{\frak r}_\eps \left(\frac{t}{\eps}\right), \delta{\frak
  p}_\eps \left(\frac{t}{\eps}\right)\right)+\eps
\sum_x\bar{\frak e}_x^{(\eps)}\left(\frac{t}{\eps}\right) \nonumber\\
&&
\\
&&
\ge \eps\sum_x\bbE_\eps{\frak e}_x\left(\delta{\frak r}_\eps \left(\frac{t}{\eps}\right), \delta{\frak
  p}_\eps \left(\frac{t}{\eps}\right)\right)+\eps
\sum_x\bar
  e_{\rm ph}(t,\eps x)-\eps
\sum_x\left|\bar{\frak e}_x^{(\eps)}\left(\frac{t}{\eps}\right)-\bar
  e_{\rm ph}(t,\eps x)\right|\nonumber
\end{eqnarray}
Letting $\eps\to0+$
and using assumptions \eqref{010702} and \eqref{010702a}  we conclude
that the limit of the energy functional appearing on the the utmost left hand side equals
\begin{equation}
\label{032603}
\int_{\bbR}e_{\rm ph}(0,y)dy\equiv \int_{\bbR}e_{\rm
  ph}(t,y)dy,\quad t\in\bbR,
\end{equation}
with $e_{\rm ph}(t,y)$ given by \eqref{bar-e}. 
Using the above together with \eqref{020506} we conclude that
$$
\int_{\bbR}e_{\rm ph}(0,y)dy\ge 
\limsup_{\eps\to0+}\eps\sum_x\bbE_\eps{\frak e}_x\left(\delta{\frak r}_\eps\left(\frac{t}{\eps}\right), \delta{\frak
  p}_\eps\left(\frac{t}{\eps}\right)\right)+
\int_{\bbR}e_{\rm ph}(t,y)dy.
$$
From  \eqref{032603}  we infer therefore  that
\begin{equation}
\label{062603}
\limsup_{\eps\to0+}\eps\sum_x\bbE_\eps{\frak e}_x\left(\delta{\frak r}_\eps\left(\frac{t}{\eps}\right), \delta{\frak
  p}_\eps\left(\frac{t}{\eps}\right)\right)=0.
\end{equation}

{
From \eqref{equiv1} applied to the configuration
$(\delta{\frak r}_\eps\left(t\eps^{-1}\right),\delta{\frak
  p}_\eps\left(t\eps^{-1}\right))$ we conclude that there exists
$c_*>0$ such that
\begin{eqnarray}
\label{equiv1a}
 &&
c_-\eps\sum_x \bbE_\eps\left[\left(\delta{\frak
      r}_{\eps,x}\left(\frac{t}{\eps}\right)\right)^2
  +\left(\delta{\frak
      p}_{\eps,x}\left(\frac{t}{\eps}\right)\right)^2\right]\nonumber\\
&&
\le 
 \eps\sum_x\bbE_\eps{\frak e}_x\left(\delta{\frak r}_\eps\left(\frac{t}{\eps}\right), \delta{\frak
  p}_\eps\left(\frac{t}{\eps}\right)\right).
\end{eqnarray}
Equalities  \eqref{020702} then follow straightforwardly from
\eqref{equiv1a} and Corollary \ref{thm1a}.}
%
%
%
\qed

\subsection{Evolution of energy density at the superdiffusive scale}

 In this section we show that the energy density disperses to infinity
 at time scale $t/\eps^{\delta}$, where $\delta\in(1,2)$.
\begin{thm}
\label{thm1aa}
Suppose that \eqref{010702} and \eqref{010702a} are in force, and $\delta\in(1,2)$.
Then,
\begin{equation}
\label{010506}
\lim_{\eps\to0+}\eps\sum_xJ(\eps x) \bbE_{\eps}{\frak e}_{x}\left(\frac{t}{\eps^{\delta}}\right)=0,
\end{equation}
for any $J\in C_0^\infty(\bbR)$, $t\ge0$. 
\end{thm}
\proof
In analogy to \eqref{070506} we write
\begin{eqnarray}
\label{070506a}
&&
\eps\sum_x\left\langle{\frak e}_x({\frak r}, {\frak
  p})\right\rangle_{\mu_\eps}=\eps\sum_x\bbE_\eps{\frak e}_x\left({\frak r}\left(\frac{t}{\eps^{\delta}}\right), {\frak
  p}\left(\frac{t}{\eps^{\delta}}\right)\right) \nonumber\\
&&
\ge \eps\sum_x\bbE_\eps{\frak e}_x\left(\delta{\frak r}_\eps \left(\frac{t}{\eps^{\delta}}\right), \delta{\frak
  p}_\eps \left(\frac{t}{\ep^{\delta}}\right)\right)\nonumber\\
&&
\\
&&
+\eps
\sum_x\bar
  e_{\rm ph}\left(\frac{t}{\eps^{\delta-1}},\eps x\right)-\eps
\sum_x\left|\bar{\frak e}_x^{(\eps)}\left(\frac{t}{\eps^{\delta}}\right)-\bar
  e_{\rm ph}\left(\frac{t}{\eps^{\delta-1}},\eps x\right)\right|.\nonumber
\end{eqnarray}
Therefore taking the limit, as $\eps\to0+$, in \eqref{070506a} we
obtain (using \eqref{080506} and \eqref{020506})
\begin{eqnarray}
\label{070506b}
&&
\int_{\bbR}e_{\rm ph}(0,y)dy
\ge \limsup_{\eps\to0+}\eps\sum_x\bbE_\eps{\frak e}_x\left(\delta{\frak r}_\eps \left(\frac{t}{\eps^{\delta}}\right), \delta{\frak
  p}_\eps \left(\frac{t}{\ep^{\delta}}\right)\right)\nonumber\\
&&
\\
&&
+\limsup_{\eps\to0+}\int_\bbR e_{\rm ph}\left(\frac{t}{\eps^{\delta-1}},y\right)dy.\nonumber
\end{eqnarray}
By \eqref{032603} we conclude 
\begin{equation}
\label{090506}
\limsup_{\eps\to0+}\eps\sum_x\bbE_\eps{\frak e}_x\left(\delta{\frak r}_\eps \left(\frac{t}{\eps^{\delta}}\right), \delta{\frak
  p}_\eps \left(\frac{t}{\ep^{\delta}}\right)\right)=0.
\end{equation}
Since, for any $J\in C_0^\infty(\bbR)$ we have
\begin{eqnarray}
\label{070506c}
&&
\eps\sum_xJ(\eps x)\bbE_\eps{\frak e}_x\left({\frak r}\left(\frac{t}{\eps^{\delta}}\right), {\frak
  p}\left(\frac{t}{\eps^{\delta}}\right)\right)\nonumber\\
&&
\\
&&
=
\eps\sum_xJ(\eps x)\bbE_\eps{\frak e}_x\left(\delta{\frak r}_\eps \left(\frac{t}{\eps^{\delta}}\right), \delta{\frak
  p}_\eps \left(\frac{t}{\eps^{\delta}}\right)\right)+\eps
\sum_xJ(\eps x)\bar{\frak e}_x^{(\eps)}\left(\frac{t}{\eps^{\delta}}\right). \nonumber
\end{eqnarray}
The conclusion of the theorem follows directly from  \eqref{090506}
and \eqref{010402-15}.
\qed

\section{Evolution of a thermal ensemble}

\label{sec4}

In this section we investigate the limit of dynamics of  ensemble families whose initial laws $(\mu_\eps)_{\eps>0}$  satisfy condition  \eqref{finite-energy1}. 
Then, as it turns out, the respective macroscopic profile of $({\frak r},{\frak p})$
is trivial, i.e. $r(y)\equiv0$ and $p(y)\equiv 0$.
In fact, we show that condition  \eqref{finite-energy1} persists in
time
and the temperature profile remains stationary at the hyperbolic time scale.


\bigskip

\subsection{Wave function}

\label{sec3.6}

Define the wave function of a given configuration
as 
\begin{equation}
\label{022302}
\psi_x:=\left(\tilde \om *{\frak q}\right)_x+i{\frak p}_x=\sum_{x'}\tilde\om_{x-x'}{\frak q}_{0,x'}+i{\frak p}_x,\quad x\in\bbZ,
\end{equation}
where ${\frak q}_{0,x'}$ is given by  \eqref{diff-q} and $(\tilde \om_x)$ are the Fourier coefficients of the dispersion
relation defined as, see \eqref{om2b}, 
\begin{equation}
\label{om2a}
\om(k):=\hat
 \al^{1/2}(k)=2 \sqrt{\tau_1}|{\frak s}( k)|\varphi({\frak s}^2( k)).
\end{equation}
Using \eqref{om2a} we can rewrite \eqref{022302} in the form
$$
\psi_x=\left(\tilde \om^{(1)} *{\frak r}\right)_x+i{\frak p}_x,\quad x\in\bbZ,
$$
with $(\tilde \om^{(1)}_x)$ are the Fourier coefficients of 
\begin{equation}
\label{om2}
\om^{(1)}(k)=-ie^{\pi i k}{\rm sign}(k)\sqrt{\tau_1}\varphi({\frak s}^2(
k)),\quad k\in\bbT.
\end{equation}
The Fourier transform $\hat \psi(k)$ of the wave function $(\psi_x)$ can be written as 
\begin{equation}
\label{h-q}
\hat\psi(k)=\om^{(1)}(k)\hat {\frak r}(k)+i \hat {\frak p}(k).
\end{equation}
Therefore,
\begin{equation}
\label{010306}
\langle
  |\hat\psi(k)|^2\rangle_{\mu_{\eps}}
=\langle|\om^{(1)}(k)\hat {\frak r}(k)|^2+| \hat {\frak p}(k)|^2\rangle_{\mu_{\eps}}+2 \langle {\rm Im}\left(\hat {\frak r}(k) \om^{(1)}(k)\hat {\frak p}^*(k)\right)\rangle_{\mu_{\eps}}.
\end{equation}
On the other hand, since $\hat {\frak r}(-k)=\hat {\frak r}^*(k)$ and
$\hat {\frak p}(-k)=\hat {\frak p}^*(k)$ we obtain
\begin{equation}
\label{020306}
\langle
  |\hat\psi(-k)|^2\rangle_{\mu_{\eps}}
=\langle|\om^{(1)}(k)\hat {\frak r}(k)|^2+| \hat {\frak p}(k)|^2\rangle_{\mu_{\eps}}-2 \langle{\rm Im}\left(\hat {\frak r}(k) \om^{(1)}(k)\hat {\frak p}^*(k)\right)\rangle_{\mu_{\eps}}.
\end{equation}
From \eqref{010306} and \eqref{020306} we obtain
\begin{equation}
\label{012302}
\int_{\bbT}{\frak  w}_\eps^2(k)dk= \int_{\bbT}\langle
  |\hat\psi(k)|^2\rangle_{\mu_{\eps}}^2dk.
\end{equation}
Condition \eqref{finite-energy1} therefore
is equivalent with
\begin{equation}
\label{finite-energy2a}
\sup_{\eps\in(0,1]}\eps^2\int_{\bbT}\langle|\hat\psi(k)|^2\rangle_{\mu_\eps}^2dk<+\infty.
\end{equation}


\bigskip

{\bf Remark.}
Note that the macroscopic profile $(r(y),p(y))$ corresponding to a given configuration $({\frak
  r},{\frak p})$ satisfying \eqref{finite-energy1} necessarily
vanishes, i.e.
\begin{equation}
  \label{eq:centered3}
 \lim_{\eps\to 0+} \eps \sum_x \langle
  {\frak r}_x\rangle_{\mu_{\eps}} J(\eps x)=0\quad\mbox{and}\quad \lim_{\eps\to 0+} \eps \sum_x \langle
  {\frak p}_x\rangle_{\mu_{\eps}} J(\eps x)=0
\end{equation}
for any $J\in C_0^\infty(\bbR)$.
To show \eqref{eq:centered3} it suffices only to prove
that 
\begin{equation}
  \label{eq:centered2}
 \lim_{\eps\to 0+} \eps \sum_x \langle
  \psi_x\rangle_{\mu_{\eps}} J(\eps x)=0.
\end{equation}
Indeed, from conditions \eqref{010702} and \eqref{010702a}, we obtain
\begin{equation}
  \label{eq:centered}
 \lim_{\eps\to 0+} \eps \sum_x \langle
  \psi_x\rangle_{\mu_{\eps}} J(\eps x)=\int_{\bbR}\psi(y)J(y)dy,
\end{equation}
where
$$
\psi(y)=\sqrt{\tau_1}{\cal H}(r)(y)+i{p}(y)
$$
and 
$$
{\cal H}(f)(y):=-\frac{1}{\pi}{\rm
  p.v.}\int_{\bbR}\frac{f(y')dy'}{y-y'}
$$
is the Hilbert transform of a given function $ f\in
C_0^\infty(\bbR)$. The integral on the right hand side is understood
in the sense of the principal value. In particular, $\psi(y)\equiv 0$ implies that
both $p(y)\equiv 0$ and $r(y)\equiv0$.

To show \eqref{eq:centered2} observe that by the Plancherel identity we can write that the
absolute value of the expression under the limit in \eqref{eq:centered2} equals
\begin{equation}
  \label{eq:centered1}
\eps\left| \int_{\bbT} \langle
  \hat\psi(k)\rangle_{\mu_{\eps}} \hat J_\eps(k)dk\right|,
\end{equation}
where
$$
\hat J_\eps(k):=\sum_x J(\eps x)\exp\left\{-2\pi i k x\right\}\approx \frac{1}{\eps}\hat J\left(\frac{k}{\eps}\right)
$$
and $\hat J(k)$ is the Fourier transform of $J(x)$. Expression in
\eqref{eq:centered1} is therefore estimated as follows
$$
\left|\int_{\bbT} \langle
  \hat\psi(k)\rangle_{\mu_{\eps}} \hat
  J\left(\frac{k}{\eps}\right)dk\right|\le \left[\int_{\bbT} \langle
  |\hat\psi(k)|^2\rangle_{\mu_{\eps}}^2 dk\right]^{1/4}\left[\int_{\bbT} \left|\hat
  J\left(\frac{k}{\eps}\right)\right|^{4/3}dk\right]^{3/4},
$$
where the estimate follows by H\"older inequality. Using
the change of variables $k':=k/\eps$ in the second integral on the
right hand side we conclude that it is bounded by
$\eps K_1^{1/4}\|\hat J\|_{L^{4/3}(\bbR)}$ for $\eps\in(0,1]$, which
proves \eqref{eq:centered2}.

\bigskip

\subsection{Evolution of the energy functional at the hyperbolic time scale}
Define by ${\frak w}_\eps(t,k)$ the energy spectrum
\eqref{022403} corresponding to the configuration $({\frak r}(t),{\frak p}(t))$.
The following result asserts that  there is  no evolution of the macroscopic temperature profile at the hyperbolic scale. Moreover, if the initial energy distribution is of the thermal type it remains so at this time scale.
\begin{thm}
\label{thm011002}
For any $J\in C_0^\infty([0,+\infty)\times \bbR)$ we have
\begin{equation}
\label{011002-c}
\lim_{\eps\to0+}\eps\sum_x \int_0^{+\infty}J(t,\eps x) \bbE_\eps {\frak e}_x\left(\frac{t}{\eps}\right)dt= \int_0^{+\infty}\int_{\bbR}J(t,y)T(y)dtdy.
\end{equation}
Moreover, at any time $t>0$ the energy spectrum  satisfies
\begin{equation}
\label{finite-energy1t}
K_1(t):=\sup_{\eps\in(0,1]}\eps^2\int_{\bbT}{\frak  w}_{\eps}^2\left(\frac{t}{\eps},k\right)dk<+\infty
\end{equation}
and
\begin{equation}
  \label{eq:centered4}
 \lim_{\eps\to 0+} \eps \sum_x 
  \bbE_\eps{\frak r}\left(\frac{t}{\eps}\right) J(\eps x)=0\quad\mbox{and}\quad  \lim_{\eps\to 0+} \eps \sum_x  \bbE_\eps{\frak p}\left(\frac{t}{\eps}\right) J(\eps x)=0
\end{equation}
for any $J\in C_0^\infty(\bbR)$.
\end{thm}
The proof of this result is given in Section \ref{sec9.4}.




\subsection{Wigner transform}

Define
$\psi_x^{(\eps)}\left(t\right):=\psi_x\left(t\eps^{-1}\right)$ and
$\hat\psi^{(\eps)}(t,k)$ its Fourier transform (belonging to $L^2(\bbT)$).
By $ W_{\eps}(t)$ we denote the (averaged) Wigner transform of
$\psi^{(\eps)}(t)$, see (5.9) of \cite{JKO},
given by
   \begin{equation}
 \label{wigner-t}
\langle W_\eps(t),J\rangle:= \int_{\bbR\times\bbT}
\widehat W_{\eps}(t,p,k)
\hat  J^*(p,k) dp dk,
\end{equation}
where   $\widehat W_{\eps}(t,p,k)$
the Fourier - Wigner transform of  $\hat \psi^{(\eps)}(t)$ given by
\begin{equation}
\label{011507}
\widehat W_{\eps}(t,p,k):=\frac{\eps}{2}\bbE_\eps\left[\left(\hat \psi^{(\eps)}\right)^*\left(t,k-\frac{\eps p}{2}\right)\hat \psi^{(\eps)}\left(t,k+\frac{\eps p}{2}\right)\right]
\end{equation}
and $J$ belongs to 
${\cal S}$--  the set of  functions on $\bbR\times
\bbT$ that are of $C^\infty$ class and such that for any
integers $l,m,n$ we have 
$$
\sup_{y\in\bbR,\,k\in\bbT} (1+y^2)^{n}|\partial_y^l\partial_k^mJ(y,k)|<+\infty.
$$ 
Let ${\cal A}$ be the completion of ${\cal S}$ under the norm
\begin{equation}
\label{norm-ta01}
\| J\|_{{\cal A}}:=\int_{\bbR}\sup_k |\hat J(p,k)|dp.
\end{equation}

In what follows we shall also consider the Fourier - Wigner anti-transform of  $\hat \psi^{(\eps)}(t)$ given by
\begin{equation}
\label{011507a}
\widehat Y_{\eps}(t,p,k):=\frac{\eps}{2}\bbE_\eps\left[\hat \psi^{(\eps)}\left(t,-k+\frac{\eps p}{2}\right)\hat \psi^{(\eps)}\left(t,k+\frac{\eps p}{2}\right)\right]
\end{equation}
From the Cauchy-Schwartz inequality  we  get
$$
|\langle W_\eps(t),J\rangle|\le \frac{\eps}{2}\|J\|_{\cal A}\bbE_\eps\|\hat
  \psi^{(\eps)}(t)\|_{L^2(\bbT)}^2.
$$
Thanks to energy conservation property of the dynamics and
Cauchy-Schwartz inequality we get
   \begin{equation}
 \label{A1}
\sup_{\eps\in(0,1]}\sup_{t\ge0}(\| Y_\eps(t)\|_{{\cal A}'}+\|
W_\eps(t)\|_{{\cal A}'})\le 2K_0,
\end{equation}
where $K_0$ is the constant appearing in condition
\eqref{finite-energy10}. 
As a direct consequence of the above estimate we infer that the family $\left(W_\eps(\cdot)\right)_{\eps\in(0,1]}$ is 
 $*-$weakly sequentially compact as $\eps\to0+$ in any
 $L^\infty([0,T];{\cal A}')$, where $T>0$,  i.e. for any $\eps_n\to0+$
 one can choose a subsequence $\eps_{n'}$ such that
 $\left(W_{\eps_{n'}}(\cdot)\right)_{n'}$ that is $*$-weakly converging. 
 In fact, using hypothesis \eqref{finite-energy1} one can prove that, see Proposition 9.1 of \cite{JKO}, the following estimate holds.
\begin{prop}
\label{prop1}
For any $M>0$ there exists $C_1>0$ such that
 \begin{eqnarray}
 \label{021501}
 &&
 \|\widehat W_{\eps}(t,p,\cdot)\|_{L^2(\bbT)}^2+\|\widehat Y_{\eps}(t,p,\cdot)\|_{L^2(\bbT)}^2\\
 &&
 \le \left(\|\widehat W_{\eps}(0,p,\cdot)\|_{L^2(\bbT)}^2+\|\widehat Y_{\eps}(0,p,\cdot)\|_{L^2(\bbT)}^2\right)e^{C_1\eps   t},\quad\forall\,\eps\in(0,1],\,|p|\le M,\,t\ge0.\nonumber
 \end{eqnarray}
\end{prop}
The right hand side of \eqref{021501} remains
bounded for $\eps\in(0,1]$ thanks to  \eqref{finite-energy2a}.

In fact asymptotically, as $\eps\to0+$, the function  $|\psi_x^{(\eps)}\left(t\right)|^2$ coincides
with the energy density ${\frak
    e}_x\left(t\eps^{-1}\right)$, which can be concluded
  from  the following result, see Proposition 5.3 of  \cite{JKO}.
\begin{prop}
\label{prop011404} Suppose that 
condition \eqref{finite-energy2a} holds.
  Then,
\begin{equation}
\label{011803}
\lim_{\eps\to0+}\eps\sum_{x}J(\eps x)  \bbE_{\eps}\left[{\frak
    e}_x\left(\frac{t}{\eps}\right)
  -\frac12\left|\psi_x^{(\eps)}\left(t\right)\right|^2\right]=0,\quad t\ge0,\,J\in C_0^\infty(\bbR).
\end{equation}
\end{prop}
From the above result we conclude that  for a real valued function
$J(y,k)\equiv J(y)$ we can write
  \begin{eqnarray}
 \label{wigner-ta}
&&
\lim_{\eps\to0+}\langle W_\eps(t),J\rangle=\lim_{\eps\to0+}\frac{\eps}{2} \sum_x
  J(\eps x) |\psi_x^{(\eps)}\left(t\right)|^2\nonumber\\
&&\\
&&
=\lim_{\eps\to0+}\eps\sum_{x}J(\eps x)  \bbE_{\eps}{\frak
    e}_x\left(\frac{t}{\eps}\right).\nonumber
\end{eqnarray}

\subsection{Homogenization of the Wigner transform in the
  mode frequency domain}

It turns out that in the limit, as $\eps\to0+$, the Wigner transform
$W_\eps(t,x,k)$ becomes independent of the $k$ variable for any
$t>0$. To avoid  boundary layer considerations at the initial time
$t=0$   we formulate this property for  
 the Laplace-Fourier transform of the respective  Wigner
function. More precisely, let
 \begin{eqnarray}
\label{lafo}
&&
\bar w_{\eps,\pm}(\la,p,k):=\int_0^{+\infty} e^{-\la t}\widehat W_{\eps}(t,p,\pm k)dt,\\
&&
\nonumber\\
&&
\bar u_{\eps,\pm}(\la,p,k):=\int_0^{+\infty} e^{-\la t}\widehat U_{\eps,\pm}(t,p,k)dt,\nonumber
\end{eqnarray}
 where
\begin{eqnarray*}
&&
\widehat U_{\eps,+}(t,p,k):=\frac{1}{2}\left[\widehat Y_{\eps}(t,p, k)+\widehat Y^*_{\eps}(t,-p, k)\right],\\
&&
\widehat U_{\eps,-}(t,p,k):=\frac{1}{2i}\left[\widehat Y_{\eps}(t,p, k)-\widehat Y^*_{\eps}(t,-p, k)\right].
\end{eqnarray*}
Thanks to 
 Proposition \ref{prop1} the 
 Laplace transform is defined for any $\la>0$ and
  for any $M>0$, compact interval $I\subset (0,+\infty)$. In addition, we have
  \begin{equation}
 \label{C-I}
 C_I:=\sup_{\eps\in(0,1]}\sup_{\la\in I,|p|\le M}\left(\|\bar w_{\eps,\iota}(\la,p)\|_{L^2(\bbT)}+\sum_{\iota\in\{-,+\}}\|\bar u_{\eps,\iota}(\la,p)\|_{L^2(\bbT)}\right)<+\infty.
 \end{equation}

Let
$$
 w_{\eps}^{(\pm)}(\la,p):=\langle \bar w_{\eps,+}(\la),\frak{e}_\pm\rangle_{L^2(\bbT)} =\langle \bar w_{\eps,-}(\la),\frak{e}_\pm\rangle_{L^2(\bbT)} ,
$$
 where 
\begin{equation}
\label{frak-e}
\frak{e}_+(k):=\frac{8}{3}{\frak s}^4( k),\quad
\frak{e}_{-}(k):=2{\frak s}^2(2 k).
\end{equation}
The following result holds, see 
 Theorem 10.2 of \cite{JKO}.
\begin{thm}
\label{cor011811a} 
Suppose that the initial laws  satisfy \eqref{finite-energy1}. Then,
for any $M>0$ and   a compact interval $I\subset (\la_0,+\infty)$ we have
\begin{equation}
\label{031811c}
\lim_{\eps\to0+}\sup_{\la\in I,|p|\le M}\int_{\bbT} \left|\bar w_{\eps,+}(\la,p,k)-
  w_{\eps}^{(\pm)}(\la,p)\right| dk=0,
\end{equation}
and
\begin{equation}
\label{031811d}
\lim_{\eps\to0+}\sup_{\la\in I,|p|\le M} \int_{\bbT} \left| \bar
  u_{\eps,\pm}(\la,p,k)\right| dk = 0.
\end{equation}
\end{thm}

\subsection{Proof of Theorem \ref{thm011002}}

\label{sec9.4}

Applying \eqref{021501} at $p=0$ we conclude that
\begin{equation}
\label{030306}
\sup_{\eps\in(0,1]}\eps^2\int_{\bbT}\left(\bbE_\eps|\hat\psi^{(\eps)}(t,k)|^2\right)^2dk<+\infty.
\end{equation}
Condition \eqref{finite-energy1t} then follows directly from \eqref{012302}.
In addition equalities \eqref{eq:centered4} can be inferred from \eqref{030306}
and \eqref{eq:centered3}.

{Concerning the proof of the convergence of the energy functional in
\eqref{011002-c}, recall that we already know that $(W_\eps(\cdot))$ is $*-$weakly
sequentially compact in  $\left(L^1([0,+\infty),{\cal A})\right)^*$. Therefore for any $\eps_n\to 0$, as
$n\to+\infty$, we can choose a subsequence, denoted in the same way,
such that  $*-$weakly 
converging to some $W \in \left(L^1([0,+\infty),{\cal A})\right)^*$.}

{
In light of \eqref{A1} we have
\begin{equation}
\label{W-Y1}
\sup_{t\ge0}\| W_\eps(t)\|_{{\cal A}'}\le K_0,
\end{equation}
with $K_0$ the same as in \eqref{finite-energy10}.
Therefore,  the respective Laplace-Fourier transform $\bar w_\eps(\la,p,k)$
can be defined for any $\la>0$. 
To identify $W$ it suffices to identify the $*$-weak limit in ${\cal
  A}'$ of the Laplace transforms $\bar w_{\eps_n}(\la)$, as $n\to+\infty$.
Thanks to Theorem \ref{cor011811a}, any limit $w(\la,p,k)\equiv w(\la,p)$
obtained in this way will be constant in $k$. In fact 
$$
w(\la,p)=\lim_{n\to+\infty} w_{\eps_n}^{(\pm)}(\la,p).
$$
We claim that in fact for any $J\in C_0(\bbR)$ 
\begin{equation}
\label{fund1}
\langle w(\la),J\rangle= \int_{\bbR}J(y)T(y)dy,
\end{equation}
which implies that in fact $W \in L^\infty([0,+\infty),{\cal A}')$
and 
$$
W(t,y,k)\equiv T(y), \quad t\ge0.
$$
In light of Proposition \ref{prop011404} this allows us to claim
\eqref{011002-c}. The only thing yet to be shown is therefore \eqref{fund1}.}

\subsubsection*{Proof of \eqref{fund1}} 
Letting $\widehat{W}_{\eps,\pm}^{(0)}=\widehat{W}_{\eps,\pm}(0)$ and $\widehat{U}_{\eps,\pm}^{(0)}=\widehat{U}_{\eps,\pm}(0)$ we obtain that for any $\la>0$ (see (10.5) of \cite{JKO})
 \begin{eqnarray}
\label{exp-wigner-eqt-1k}
&&\la \bar w_{\eps,+}-\widehat W_{\eps,+}^{(0)}=-i\delta_\eps\om
\bar w_{\eps,+}-i\ga R'p\bar u_{\eps,-}
 +\frac{\ga}{\eps}{\cal L}
\left(\bar w _{\eps,+}-\bar u_{\eps,+}\right)
+\eps\bar r^{(1)}_\eps,\nonumber
\\
&&
\\
&&
\la \bar u_{\eps,+}-\widehat U_{\eps,+}^{(0)}=\frac{2\bar\om}{\eps}
\bar u_{\eps,-}
+\frac{\ga}{\eps}{\cal  L}\left[\bar u_{\eps,+}-\frac{1}{2}\left(\bar w_{\eps,+}+\bar w_{\eps,-}\right)\right]
+\eps\bar r^{(2)}_\eps,
\nonumber
\\
&&
\nonumber\\
&&
\la \bar u_{\eps,-}-\widehat U_{\eps,-}^{(0)}=-\frac{2\bar\om}{\eps}
\bar u_{\eps,+}
-\frac{2\ga}{\eps}R\bar u_{\eps,-}
-\frac{i\ga R'p}{2}(\bar
w_{\eps,-}-\bar w_{\eps,+})
 +\eps\bar r^{(3)}_\eps
\nonumber\\
&&
\nonumber\\
&&\la \bar w_{\eps,-}-\widehat W_{\eps,-}^{(0)}=i\delta_\eps\om
\bar w_{\eps,-}+i\ga R'p\bar u_{\eps,-}
+\frac{\ga}{\eps}{\cal L}\left(\bar w_{\eps,-}-\bar u_{\eps,+}\right)
+\eps\bar r^{(4)}_\eps,\nonumber
\end{eqnarray}
where 
\begin{eqnarray}
\label{scat}
&&
\delta_{\eps}\om (k,p):=\frac{1}{\eps}\left[\om\left(k+\frac{\eps p}{2}\right)-\om\left(k-\frac{\eps p}{2}\right)\right],\nonumber\\
&&
\\
&&
\bar \om(k,p):=\frac{1}{2}\left[\om\left(k+\frac{p}{2}\right)+\om\left(k-\frac{ p}{2}\right)\right],\nonumber\\
&&
 {\cal L} w(k):=2\int_{\bbT} R(k,k') w(k') dk'-2R(k)w(k),\quad w\in L^1( \bbT),\nonumber
 \end{eqnarray}
and
  \begin{eqnarray}
 \label{060411}
&&
R(k,k')
:=
\frac{3}{4}\sum_{\iota\in\{-,+\}} \frak e_\iota(k)  \frak
e_{-\iota}(k'),\\
&&
R(k):=\int_{\bbT}R(k,k')dk',\nonumber
\end{eqnarray}
with $\frak{e}_\pm(k)$ given by \eqref{frak-e}.
The remainder terms satisfy
 \begin{equation}
 \label{R-I}
\sup_{\eps\in(0,1]}\sup_{\la\in I,|p|\le M}\sum_{j=1}^4\|\bar r_{\eps}^{(j)}(\la,p)\|_{L^2(\bbT)}<+\infty
 \end{equation}
  for any $M>0$, compact interval $I\subset (0,+\infty)$.

From the first equation of the system \eqref{exp-wigner-eqt-1k} we get
\begin{equation}
\label{020911a}
 D^{(\eps)}\bar w_{\eps} 
=\eps\widehat W_{\eps}^{(0)} +\frac 32\ga\sum_{\iota\in\{-,+\}}\frak
e_{\iota} w_\eps^{(-\iota)}+q_\eps,
\end{equation}
where 
\begin{equation}
\label{030911}
 D^{(\eps)}:= \eps \la + 2\ga
R + i\eps\delta_\eps\om,\\
\end{equation}
and $q_\eps:=\sum_{i=1}^3 q_\eps^{(i)}$, with
\begin{equation}
q_\eps^{(1)}
:=\eps^{2}\bar r^{(1)}_\eps,\quad
q_\eps^{(2)}:=-\ga{\cal L}\bar u_{\eps,+},\quad
q_\eps^{(3)}
:=
-i\eps\ga R'p\bar u_{\eps,-}.
\end{equation}
Computing $\bar w_{\eps}$  from \eqref{020911a} and then multiplying
scalarly both sides of the resulting equation by $\ga\frak e_\iota$,
$\iota\in\{-,+\}$ we get the following system 
\begin{eqnarray*}
&&
\ga w_{\eps}^{(\iota)}\int_{\bbT}\left(1-\frac{3\ga \frak e_-\frak
    e_+}{2D^{(\eps)}}\right) dk -\frac{3\ga^2}{2}
 w_{\eps}^{(-\iota)}\int_{\bbT}\frac{ \frak e_{\iota}^2}{D^{(\eps)}} dk\\
&&
=\ga\eps\int_{\bbT}\frac{ \frak e_{\iota} 
\widehat W_{\eps}^{(0)} }{D^{(\eps)}} dk
+\ga\int_{\bbT} 
\frac{ \frak e_{\iota} q_\eps }{D^{(\eps)}} dk,\quad\iota\in\{-,+\}.
\end{eqnarray*}
Adding sideways the above equations corresponding to both values of
$\iota$ and then dividing both sides of the resulting equation by $\eps$ we obtain
\begin{equation}
\label{032301}
  a_w^{(\eps)} w_{\eps}^{(+)} -a_+^{(\eps)}\left(
    w_{\eps}^{(+)} - w_{\eps}^{(-)}\right)
  =\frac{4\ga}{3}\int_{\bbT} \frac{R\widehat
    W_{\eps}^{(0)}}{D^{(\eps)}} dk
  +\frac{4\ga}{3\eps}\int_{\bbT}
  \frac{R q_\eps}{D^{(\eps)}} dk,
\end{equation}
where
\begin{eqnarray}
\label{a-w}
&&
a_w^{(\eps)}(\lambda, p) :=
\frac{4\ga}{3} \int_{\bbT}
\frac{ \la + i\delta_\eps\om}{D^{(\eps)}} R \;dk\\
&&
a_+^{(\eps)}(\lambda,p):= 
\ga
\int_{\bbT}\frac{ \la + i\delta_\eps\om}{D^{(\eps)}} 
\frak e_+ \; dk. \nonumber
\end{eqnarray}
Equality \eqref{fund1} is then a consequence of the following.
\begin{prop}
\label{cor012301}
{For any $J\in {\cal S}$ such that $J(y,k)\equiv J(y)$
  and $\la>\la_0$ we have
\begin{equation}
\label{011903}
\lim_{\eps\to0+}
\left(\int_{\bbR\times \bbT} 
\frac{2\ga R}{D^{(\eps)}}\,
 \widehat W_{\eps}^{(0)}\hat Jdpdk- \int_{\bbR\times \bbT} \hat T \hat J dpdk \right)=0,
\end{equation}}
In addition, for any $M>0$ and a compact interval $I\subset(0,+\infty)$
\begin{equation}
\label{052301}
\lim_{\eps\to0+}\sup_{\la\in I,|p|\le M} \left|a_+^{(\eps)}(\la,p)
  \left( w_{\eps}^{(+)}(\la,p) - w_{\eps}^{(-)}(\la,p)\right)\right|=0,
\end{equation}
\begin{equation}
\label{010301}
\lim_{\eps\to0+}\sup_{\la\in I,|p|\le M}\left|a_w^{(\eps)}(\la,p)-\frac{2\la}{3}\right|=0,
\end{equation}
and
\begin{equation}
\label{021903}
\lim_{\eps\to0+}\sup_{\la\in I,|p|\le
  M}\left|\frac{4\ga}{3\eps}\int_{\bbT}\frac{R q_\eps
  }{D^{(\eps)}} dk
  \right|=0.
\end{equation}
\end{prop}
\proof
Equality \eqref{011903} follows from \eqref{ini} and \eqref{wigner-ta}. Formulas \eqref{052301} and \eqref{010301} can be easily substantiated using the Lebesgue dominated convergence theorem. To argue \eqref{021903} it suffices only to prove that 
\begin{equation}
\label{021903j}
\lim_{\eps\to0+}\sup_{\la\in I,|p|\le
  M}\left|\frac{4\ga}{3\eps}\int_{\bbT}\frac{R q_\eps^{(j)}
  }{D^{(\eps)}} dk
  \right|=0,\quad j=2,3.
\end{equation}
Note that
\begin{eqnarray*}
\left|\frac{4\ga}{3\eps}\int_{\bbT}\frac{R q_\eps^{(3)}
  }{D^{(\eps)}} dk
  \right|\le \frac{4\ga^2|p|\|R'\|_{\infty}}{3}\int_{\bbT}\frac{R |\bar u_{\eps,-}|
  }{|D^{(\eps)}|} dk\le C\int_{\bbT}|\bar u_{\eps,-}|
 dk
\end{eqnarray*}
and \eqref{021903j} for $j=3$ follows from \eqref{031811d}. 

In the case $j=2$ note that $\int_{\bbT} q_\eps^{(2)}dk=0$, therefore
$$
\frac{4\ga}{3\eps}\int_{\bbT}\frac{R q_\eps^{(2)}
  }{D^{(\eps)}} dk
  =\frac{2}{3}\int_{\bbT}\frac{(\la + i\delta_\eps\om) q_\eps^{(2)}
  }{D^{(\eps)}} dk.
$$
Since $k\mapsto \delta_\eps\om(k,p)$ is odd for each $p$ we get that the latter integral equals
\begin{eqnarray}
\label{092301}
&&
I_\eps:=-\frac{2\ga}{3}\int_{\bbT}\frac{\la(\eps\la  + 2\ga R) +\eps(\delta_\eps\om)^2
  }{|D^{(\eps)}|^2} {\cal L}\bar u_{\eps,+}dk.\nonumber\\
&&
\\
&&
=-\frac\ga2\sum_{\iota\in\{-,+\}}u_{\eps,+}^{-\iota}\int_{\bbT}\frac{
\frak e_{\iota}  
}{|D^{(\eps)}|^2} \left[\la (\eps\la+2\ga
  R_\eps)+\eps( \delta_\eps\om)^2 \right]dk\nonumber\\
&&
\nonumber\\
&&
+\frac{4\ga}{3}\int_{\bbT}\frac{
 R \bar u_{\eps,+}
}{|D^{(\eps)}|^2} \left[\la (\eps\la+2\ga
  R_\eps)+\eps( \delta_\eps\om)^2 \right]dk,\nonumber
\end{eqnarray}
with 
$$
 u_{\eps,\iota}^{(\iota')}(\la,p):=\langle \bar u_{\eps,\iota}(\la),\frak{e}_{\iota'}\rangle_{L^2(\bbT)},\quad \iota,
 \iota'\in\{-,+\}.
$$

We conclude therefore that there exists a constant $C>0$ such that
\begin{equation}
\label{102301}
|I_\eps|\le C \ga \eps\sum_{\iota\in\{-,+\}}| u_{\eps,+}^{-\iota}|\int_{\bbT}\frac{
\frak e_{\iota} ( \delta_\eps\om)^2
}{|D^{(\eps)}|^2} dk
+\ga \eps\int_{\bbT}\frac{
 R |\bar u_{\eps,+}|
( \delta_\eps\om)^2}{|D^{(\eps)}|^2} dk.
\end{equation}
Denote the terms appearing on the right hand side by
$I_\eps^{(1)}$ and $I_\eps^{(2)}$, respectively.
Since $|D^{(\eps)}|^2\ge 2\eps\la \ga R$ we conclude that there exists $C>0$ such that
$$
I_\eps^{(1)}\le C \sum_{\iota\in\{-,+\}}|u_{\eps,+}^{\iota}|
\quad\mbox{
and}
\quad
I_\eps^{(2)}\le C\|\bar u_{\eps,+}\|_{L^1(\bbT)}.
$$
Therefore \eqref{021903j} for $j=2$ follows directly from \eqref{031811d}.
\qed


\subsection{The limit of energy functional at the superdiffusive time scale}

We have shown in Theorem \ref{thm011002} that the evolution of the
temperature profile takes place on a scale longer than the hyperbolic one.
In fact the right time-space scaling is given by
$(t\eps^{-3/2},x\eps^{-1})$ as it has been shown in 
Theorem 3.1 of   \cite{JKO}.
\begin{thm}
\label{energy-prop-main1}
 Suppose that the distribution of the initial configuration $ ({\frak r},{\frak
  p})$ satisfies 
condition \eqref{finite-energy1}.
 Then, for any test function $J\in 
C^{\infty}_0([0,+\infty)\times \bbR)$ we  have: 
\begin{equation}
\label{041803a}
 \lim_{\eps\to0+}\eps\sum_{x}\int_0^{+\infty} J(t,\eps x)  \bbE_{\eps}{\frak
   e}_{x}\left(\frac{t}{\eps^{3/2}}\right) dt
=   \int_0^{+\infty}\int_{\bbR}T(t,y) J(t,y)dtdy.
 \end{equation}
Here $T(t,y)$ satisfies the fractional heat equation:
\begin{equation}
  \label{eq:frheat}
  \partial_t T(t,y) = - \hat c|\Delta_y|^{3/4}T(t,y)
\end{equation}
with the initial condition $T(0,y)=T(y)$, given by \eqref{ini}
and 
\begin{equation}
\label{hatc-32a}
\hat c:=\frac{[\al''(0)]^{3/4}}{2^{9/4}(3\ga)^{1/2}}.
\end{equation}
\end{thm}

\section{Proof of Theorem \ref{cor011002} and Corollary \ref{cor031102}}

\label{sec-phon1}

Since the dynamics is linear, the  solutions of  \eqref{eq:bas2} are of
the form
$$
{\frak r}_x(t)={\frak r}'_x(t)+{\frak r}''_x(t)\quad \mbox{ and }\quad
{\frak p}_x(t)={\frak p}'_x(t)+{\frak p}''_x(t),\quad t\ge0,\,x\in\bbZ,
$$
where $({\frak r}'(t),{\frak p}'(t))$ and  $({\frak r}''(t),{\frak
  p}''(t))$ are the solutions corresponding to   the  initial configurations
 $({\frak r}'_x,{\frak
  p}'_x)$ and $({\frak r}_x'',{\frak
  p}_x'')$   under the dynamics \eqref{eq:bas2}, see Section
\ref{sec3a1}.
The convergence of the ${\frak r}$ and ${\frak p}$ components of the
vector ${\frak w}(t)$ from Corollary \ref{cor031102} is therefore
a direct consequence of the conclusions of Theorems \ref{thm011002} and \ref{thm1}

The statements concerning the asymptotics of the thermal and phononic
components of the energy functional contained in \eqref{011002-a} and \eqref{011002-b}
also follow from the aforementioned theorems.
To prove \eqref{011002-d},  finishing in this way also the proof of Corollary  \ref{cor031102},  note that
\begin{equation}
\label{decomp}
{\frak e}_x\left(\frac{t}{\eps}\right)={\frak e}_{{\rm th},x}\left(\frac{t}{\eps}\right)+
{\frak e}_{{\rm ph},x}\left(\frac{t}{\eps}\right)+2{\frak
  e}_x\left({\frak r}'\left(\frac{t}{\eps}\right),{\frak
    p}'\left(\frac{t}{\eps}\right);
{\frak r}''\left(\frac{t}{\eps}\right),{\frak p}''\left(\frac{t}{\eps}\right)\right)
\end{equation} 
where ${\frak e}_{{\rm th},x}(t)$ and ${\frak e}_{{\rm ph},x}(t)$
are defined in \eqref{050306}. Given two configurations
$({\frak r}^{(j)},{\frak p}^{(j)})$, $j=1,2$ the ''mixed'' energy
functional is defined as
\begin{equation}
\label{050306a}
{\frak e}_{x}({\frak r}^{(1)},{\frak p}^{(1)};{\frak r}^{(2)},{\frak
  p}^{(2)}):=
\frac{1}2 \frak p_x^{(1)} \frak p_x^{(2)} - \frac 14 \sum_{x'} \alpha_{x-x'} \frak
  q_{x,x'}^{(1)} \frak
  q_{x,x'}^{(2)}.
\end{equation}
Here $\frak
  q_{x,x'}^{(j)}$ are computed from \eqref{diff-q} for the respective
  configurations  ${\frak r}^{(j)}$, $j=1,2$.
Formula  \eqref{011002-d} is a simple consequence of the following.
\begin{lemma}
\label{lm010306}
Under the assumptions of Theorem  \ref{cor011002} we have
\begin{equation}
\label{011002-d1}
\lim_{\eps\to0+}\eps\sum_xJ(\eps x)\bbE_\eps{\frak
  e}_x\left({\frak r}'\left(\frac{t}{\eps}\right),{\frak
    p}'\left(\frac{t}{\eps}\right);
{\frak r}''\left(\frac{t}{\eps}\right),{\frak
  p}''\left(\frac{t}{\eps}\right)\right)=0
\end{equation}
for any $J\in C_0^\infty(\bbR)$ and $ t\ge0$.
\end{lemma}
\proof
Let (see \eqref{diff-dq})
$$
\delta_\eps{\frak r}''_x\left(\frac{t}{\eps}\right):={\frak
  r}''_x\left(\frac{t}{\eps}\right)-r(t,\eps x),\quad \delta_\eps{\frak
    p}''\left(\frac{t}{\eps}\right):={\frak
    p}''\left(\frac{t}{\eps}\right) -p(t,\eps x).
$$
Using Cauchy-Schwartz inequality we can estimate
\begin{eqnarray*}
&&\eps\left|\sum_x\bbE_\eps{\frak
  e}_x\left({\frak r}'\left(\frac{t}{\eps}\right),{\frak
    p}'\left(\frac{t}{\eps}\right);
\delta_\eps{\frak r}''\left(\frac{t}{\eps}\right),\delta_\eps{\frak
  p}''\left(\frac{t}{\eps}\right)\right)\right|\\
&&
\le
\left\{\eps\sum_x\bbE_\eps{\frak
  e}_x\left({\frak r}'\left(\frac{t}{\eps}\right),{\frak
    p}'\left(\frac{t}{\eps}\right)\right)\right\}^{1/2}
\left\{
\eps\sum_x\bbE_\eps{\frak
  e}_x\left(
\delta_\eps{\frak r}''\left(\frac{t}{\eps}\right),\delta_\eps{\frak
  p}''\left(\frac{t}{\eps}\right)\right)\right\}^{1/2}
\end{eqnarray*}
The first factor on the right hand side stays bounded, due to the conservation of energy property of
the dynamics, while the second one vanishes thanks to \eqref{062603}.  
We can  conclude therefore that
\begin{equation}
\label{011002-e}
\lim_{\eps\to0+}\eps\left|\sum_x\bbE_\eps{\frak
  e}_x\left({\frak r}'\left(\frac{t}{\eps}\right),{\frak
    p}'\left(\frac{t}{\eps}\right);
\delta_\eps{\frak r}''\left(\frac{t}{\eps}\right),\delta_\eps{\frak p}''\left(\frac{t}{\eps}\right)\right)\right|=0,
\end{equation}
for any $t\ge0$, 
Equality \eqref{011002-d1} would follow, provided we can show that
\begin{equation}
\label{011002-f}
\lim_{\eps\to0+}\eps\sum_xJ(\eps x) \bbE_\eps{\frak
  e}_x\left({\frak r}'\left(\frac{t}{\eps}\right),{\frak
    p}'\left(\frac{t}{\eps}\right);
r_\eps(t),p_\eps(t)\right)=0,
\end{equation}
for any $J\in C_0^\infty(\bbR)$ and $t\ge0$, where
$$
(r_\eps(t),p_\eps(t)):=\left(r(t,\eps x),p(t,\eps x)\right)_{x\in\bbZ}.
$$
The latter however is a direct consequence of
\eqref{eq:centered4}.
\qed


\section{Proof of Theorem \ref{energy-prop-main}}

\label{sec-phon}

Using the notation from Section \ref{sec-phon1}
we
can write the analogue of \eqref{decomp} at the time scale
$t/\eps^{3/2}$. 
 Thanks to \eqref{010506} for any $J\in C_0^\infty(\bbR)$ we have
\begin{equation}
\label{010608}
\lim_{\eps\to0+}\sum_xJ(\eps x)\bbE_\eps{\frak e}_{{\rm ph},x}\left(\frac{t}{\eps^{3/2}}\right)=0,
\end{equation}
The respective time-space weak limit of   ${\frak e}_{{\rm th},x}\left(t/\eps^{3/2}\right)$ can be evaluated 
using Theorem \ref{energy-prop-main}.
Finally, to finish the proof of Theorem  \ref{energy-prop-main}  we
need the following analogue of Lemma \ref{lm010306}.
\begin{lemma}
\label{lm010306a}
Under the assumptions of Theorem  \ref{energy-prop-main}   we have
\begin{equation}
\label{011002-d2}
\lim_{\eps\to0+}\eps\sum_xJ(\eps x)\bbE_\eps{\frak
  e}_x\left({\frak r}'\left(\frac{t}{\eps^{3/2}}\right),{\frak
    p}'\left(\frac{t}{\eps^{3/2}}\right);
{\frak r}''\left(\frac{t}{\eps^{3/2}}\right),{\frak
  p}''\left(\frac{t}{\eps^{3/2}}\right)\right)=0
\end{equation}
for any $J\in C_0^\infty(\bbR)$
\end{lemma}
\proof
Using Cauchy-Schwartz inequality we can estimate
\begin{eqnarray*}
&&\eps\left|\sum_xJ(\eps x)\bbE_\eps{\frak
  e}_x\left({\frak r}'\left(\frac{t}{\eps^{3/2}}\right),{\frak
    p}'\left(\frac{t}{\eps^{3/2}}\right);
{\frak r}''\left(\frac{t}{\eps^{3/2}}\right),{\frak
  p}''\left(\frac{t}{\eps^{3/2}}\right)\right)\right|\\
&&
\le
\left\{\eps\|J\|_\infty\sum_x\bbE_\eps{\frak
  e}_x\left({\frak r}'\left(\frac{t}{\eps^{3/2}}\right),{\frak
    p}'\left(\frac{t}{\eps^{3/2}}\right)\right)\right\}^{1/2}
\\
&&
\times \left\{
\eps\sum_xJ(\eps x)\bbE_\eps{\frak
  e}_x\left(
{\frak r}''\left(\frac{t}{\eps^{3/2}}\right),{\frak
  p}''\left(\frac{t}{\eps^{3/2}}\right)\right)\right\}^{1/2}
\end{eqnarray*}
The first factor on the right hand side stays bounded, due to the conservation of energy property of
the dynamics, while the second one vanishes thanks to \eqref{010608}.  
This ends the proof of the lemma.\qed

\section{Proof of Theorem \ref{p-modes}}
 \label{sec9}

From the definition of the normal modes, see \eqref{phonon-1}, we conclude that
\begin{eqnarray}
\label{011703}
&&
d\hat{\frak f}^{(\pm)}(t,k)=\left\{\vphantom{\int_0^1}\pm(1-e^{-2i\pi k})\sqrt{\tau_1}\hat{\frak  f}^{(\pm)}(t,k)\right.\\
&&
\left.\vphantom{\int_0^1}+D_\pm(1-e^{2i\pi k})^2\hat{\frak f}^{(\pm)}(t,k)\vphantom{\int_0^1}+O({\frak s}^3( k))\right\}dt+dM_t(k),\nonumber
\end{eqnarray}
where 
\begin{equation}
\label{M-t}
dM_t(k):=2i\ga^{1/2}\int_{\bbT} r(k,k')\hat{\frak p}(t,k-k')B(dt,dk'),
\end{equation} 
with $B(dt,dk)$ Gaussian white noise in $(t,k)$, satisfying
\begin{eqnarray*}
&&
B^*(dt,dk)=B(dt,-dk),\\
&&
\\
&&
 \bbE[B(dt,dk)B^*(dt',dk')]=\delta(t-t')\delta(k-k')
\end{eqnarray*}
and $D_\pm$ are given by 
\begin{equation}
\label{Cpm}
D_\pm:=\frac{3\ga\pm\sqrt{\tau_1}}{2}.
\end{equation}
Let 
$$
\hat{\bar{\frak f}}^{(\pm)}_\eps\left(t,k\right):=\eps\bbE_\eps
\hat{\frak f}^{(\pm)}\left(\frac{t}{\eps},\eps k\right),\quad k\in  \eps^{-1}\bbT.
$$
It satisfies
\begin{equation}
\label{021703}
\frac{d}{dt}\hat{\bar{\frak f}}^{(\pm)}_\eps\left(t,k\right)=\pm\frac{1}{\eps}(1-e^{-2i\eps\pi k})\hat{\bar{\frak f}}^{(\pm)}_\eps\left(0,k\right)+O(\eps).
\end{equation}
An elementary stability theory for solutions of ordinary differential equations guarantees
that for any $T,M>0$ we have
\begin{equation}
\label{031703}
\hat{\bar{\frak f}}^{(\pm)}_\eps\left(t,k\right)=e^{\pm2i\pi k t}(1+O(\eps))\hat{\bar{\frak f}}^{(\pm)}_\eps\left(t,k\right),\quad |k|\le M,\,|t|\le T.
\end{equation}
After a straightforward calculation we obtain that the left hand side of \eqref{u1} equals
$$
\lim_{\eps\to0+}\int_{\bbR}\hat J(p) \hat{\bar{\frak f}}^{(\pm)}_\eps\left(t,- p\right)dp=
\lim_{\eps\to0+}\int_{\bbR}\hat J(p)e^{\mp 2i\pi p t}  \hat{\bar{\frak f}}^{(\pm)}_{\eps}\left(0,- p\right)dp,
$$
which tends to the expression on the right hand side of \eqref{u1}, as
$\eps\to0+$ for any $J\in C_0^\infty(\bbR)$.

Let 
$$
\bar{\frak v}_\eps^{(\iota)}(t,k):=  \hat{\bar {\frak
  f}}^{(\iota)}_\eps\left(\frac{t}{\eps},
  k\right)\exp\left\{-\iota(1-e^{-2i\pi \eps k})\frac{\sqrt{\tau_1}
    t}{\eps^2}\right\},\quad \iota\in\{-,+\}
$$
for $k\in\eps^{-1}\bbT$. For any $T,M>0$  and $\iota\in\{-,+\}$ we have
\begin{equation}
\label{V+}
\frac{d\bar{\frak v}_\eps^{(\iota)}}{dt}(t,k)=\frac{D_{\iota}}{\eps^2}(1-e^{2i\pi \eps k})^2\bar{\frak v}_\eps^{(\iota)}(t,k)\vphantom{\int_0^1}+O(\eps),\,|k|\le M,\,t\in[0,T].
\end{equation}
Using again the elementary stability theory of ordinary differential
equations we conclude that 
$$
\lim_{\eps\to0+}\sup_{t\in[0,T],|k|\le M}\left|\bar{\frak
    v}_\eps^{(\iota)}(t,k)-  \hat{\bar{\frak
    f}}^{(\iota)}_\eps\left(0, k\right)e^{-4\pi^2
    tD_{\iota}|k|^2}\right|=0 ,\quad \iota\in\{-,+\}.
$$
For any $ J\in C_0^\infty(\bbR)$  we can write that the expression under the limit on the left hand side of \eqref{ud1} equals
\begin{eqnarray*}
&&
\lim_{\eps\to0+} \int_{\bbR}\hat J(p)\exp\left\{-\iota 2\pi ip\sqrt{\tau_1}\frac{t}{\eps}\right\}\hat{\bar {\frak f}}^{(\iota)}_\eps\left(\frac{t}{\eps},- p\right)dp\\
&&
=\lim_{\eps\to0+}\int_{\bbR}\hat J(p)\exp\left\{\iota\sqrt{\tau_1}\left(1-e^{-2\pi i\eps p}-2\pi i\eps p\right)\frac{t}{\eps^2}\right\}\bar{\frak v}_\eps^{(\iota)}\left(t,- p\right)dp
\\
&&
=\lim_{\eps\to0+}\int_{\bbR}\hat J(p)\exp\left\{2\iota\sqrt{\tau_1}\pi^2p^2 t+O(\eps)\right\}\bar{\frak v}_\eps^{(\iota)}\left(t,- p\right)dp
\end{eqnarray*}
and the latest limit equals
$$
\int_{\bbR}\hat J(p)\hat f^{(\iota)}(-p)e^{-4\pi^2 tD|p|^2}dp,
$$
with $\hat f^{(\iota)}(p)$ the Fourier transform of
$f^{(\iota)}(y)$, given by \eqref{012504}, which ends the proof of \eqref{ud1}.\qed

\bigskip

\section{Examples}

\label{sec10}

In the final section we 
give  examples of the initial data that are either of
thermal or phononic types introduced  in Definitions \ref{thermal}
and \ref{phononic}. The examples are formulated in terms of the wave function.

\subsection{Non-random initial data}

Suppose that  $\phi(x)$ is a function that belongs to $C_0^\infty(\bbR)$. Let $a\ge 0$ and let 
$\mu_\eps$ be $\delta$-type measures on $\ell_2$  concentrated at 
\begin{equation}
\label{basic}
\psi_x^{(\eps)}:=\eps^{(a-1)/2}\phi(\eps^a x),\quad x\in\bbZ.
\end{equation}

We have
$$
\hat\psi_\eps(k)=\eps^{(a-1)/2}\int_{\bbR}\hat\phi(p)\left[\sum_{x\in\bbZ}\exp\left\{2\pi ix ( \eps^a p- k)\right\}\right]dp.
$$
Using  Poisson summation formula  (see e.g. p. 566 of \cite{lax})
\begin{equation}
\label{poisson}
\sum_{x\in\bbZ}\exp\left\{ i b x \xi\right\}=\frac{2\pi}{|b|}\sum_{x\in\bbZ}\delta_0\left(\xi-\frac{2\pi}{b}x\right),\quad \xi\in\bbR,
\end{equation}
{(understood in the distribution sense)} that holds for any $b\not=0$,
 and the fact
that $\hat\phi(k)$ is rapidly decaying we conclude
$$
\hat\psi_\eps(k)
\approx\frac{1}{\eps^{(a+1)/2}}\hat\phi\left(\frac{k}{\eps^a}\right).
$$


\subsubsection{Case $a=1$ -  macroscopic initial data}

Note that then
\begin{equation}
\label{psi-0}
\limsup_{\eps\to0+} \int_\bbT dk\; \left[\eps\left<|\hat \psi(k)|^2
        \right>_{\mu_\eps}\right]=\int_{\bbR}|\hat\phi(p)|^2dp <+\infty,
\end{equation}
where $\hat\phi(p)$ is the Fourier transform of $\phi(y)$. The data is
of phononic type, in the sense of Definition \ref{phononic}, with the
macroscopic profile given by $\phi(y)$.
On the other hand, condition \eqref{finite-energy1} fails, as can be seen from the following computation:
$$
 \int_\bbT dk\; \left[\eps\left<|\hat \psi(k)|^2
        \right>_{\mu_\eps}\right]^2=
       \eps^2 \int_\bbT |\hat\psi_\eps(k)|^4dk
       \approx \frac{1}{\eps^2}\int_{-1/2}^{1/2} \left|\hat\phi\left(\frac{k}{\eps}\right)\right|^4dk
        \approx\frac{1}{\eps} \int_{\bbR} |\hat\phi\left(p\right)|^4dp.
$$

\subsubsection{Case $a\in(0,1)$ -  oscillating (but not too fast) data}

In this case one can easily verify that condition
\eqref{finite-energy10} holds, but again 
condition \eqref{finite-energy1} fails. However, the rate of the blow-up is slower than in the case of the macroscopic data. Indeed,
\begin{eqnarray*}
&&
 \int_\bbT dk\; \left[\eps\left<|\hat \psi(k)|^2
        \right>_{\mu_\eps}\right]^2=
       \eps^2 \int_\bbT |\hat\psi_\eps(k)|^4dk\\
        &&
        \approx\frac{1}{\eps^{a}} \int_{-1/(2\eps^a)}^{1/(2\eps^a)} |\hat\phi\left(k\right)|^4dk\approx\frac{1}{\eps^a} \int_{\bbR} |\hat\phi\left(p\right)|^4dp.
\end{eqnarray*}
On the other hand, it is easy to verify that the macroscopic profile
for the initial data
vanishes but
$$
K_0=\lim_{\eps\to0+} \eps\sum_x\left<| \psi_x^{(\eps)}|^2
        \right>_{\mu_\eps}
        = \int_{\bbR} |\hat\phi\left(k\right)|^2dk.
$$
Hence the family $(\mu_\eps)_{\eps>0}$ is neither of phononic, nor
thermal type.

\subsubsection{Case $a=0$ - microscopically  oscillatory data}

We have
$$
\hat\psi_\eps(k)=\eps^{-1/2}\sum_{x\in\bbZ}\phi( x)\exp\left\{-2\pi i  x k\right\}
=\eps^{-1/2}\tilde \phi(k),
$$
where, the periodized Fourier transform of $\phi(y)$ is given by
\begin{equation}
\label{tphi}
\tilde\phi(k)=\sum_{x\in\bbZ}\hat\phi(x+k).
\end{equation}
In this case
condition \eqref{finite-energy1} holds, since
$$
 \int_\bbT dk\; \left[\eps\left<|\hat \psi(k)|^2
        \right>_{\mu_\eps}\right]^2
               = \int_{\bbT} |\tilde\phi\left(k\right)|^4dk.
$$
The data is of  thermal type.

\subsection{Random initial data}

\subsubsection{Modified stationary field}
Assume that $(\eta_x)_{x\in\bbZ}$ is a zero mean, random
stationary field such that $\bbE|\eta_0|^2<+\infty$. We suppose that its covariance
can be written as 
\begin{equation}
\label{r-x}
r_x=\bbE\left(\eta_x\eta_0^*\right)=\int_{\bbT}\exp\left\{2\pi i k
  x\right\}\hat R(k)dk,\quad x\in\bbZ,
\end{equation}
where $\hat R\in C(\bbT)$ is  non-negative.
Given $a\ge 0$ and  
$\phi(x)\in C_0^\infty(\bbR)$ define the wave function as 
\begin{equation}
\label{basic1}
\psi_x^{(\eps)}:=\eps^{(a-1)/2}\phi(\eps^a x)\eta_x,\quad x\in\bbZ.
\end{equation}
One can easily check  that condtion \eqref{finite-energy10} holds. We
show that both micro- and
macroscopically varying initial data satisfy condition \eqref{finite-energy1}.

For  $a\in(0,1]$ (the oscillatory case) we have
\begin{equation}
\label{020104}
\hat\psi^{(\eps)}(k)\approx\frac{1}{\eps^{(a+1)/2}}\int_{\bbT}\hat\phi\left(\frac{k-\ell}{\eps^a}\right)\hat\eta(d\ell),
\end{equation}
where $\hat\eta(d\ell)$ is the stochastic spectral measure
corresponding to $(\eta_x)_{x\in\bbZ}$. Then, thanks to \eqref{020104}
we get
\begin{eqnarray}
\label{psi3}
&&
 \int_\bbT dk\; \left\{\eps\bbE|\hat \psi^{(\eps)}(k)|^2
     \right\}^2\approx\eps^{-2a}\int_{\bbT}dk \left\{\int_{\bbT}
       \left|\hat\phi\left(\frac{k-\ell}{\eps^a}\right)\right|^2\hat
       R(\ell)d\ell\right\}^2\nonumber\\
&&
\\
&&
=
\eps^{-2a}\int_{\bbT}dk \left\{\int_{\bbT^2} \left|\hat\phi\left(\frac{k-\ell}{\eps^a}\right)\right|^2\left|\hat\phi\left(\frac{k-\ell'}{\eps^a}\right)\right|^2\hat R(\ell)
\hat R(\ell')d\ell d\ell'\right\}.
\nonumber
\end{eqnarray}
Changing variables $\tilde \ell':=\ell'/\eps^a$ and $\tilde k:=k/\eps^a$ we get
that the last expression equals
$$
\int_{\bbT}\hat R(\ell)d\ell \left\{\int_{-1/(2\eps^a)}^{1/(2\eps^a)}\int_{-1/(2\eps^a)}^{1/(2\eps^a)} \left|\hat\phi\left(k-\frac{\ell}{\eps^a}\right)\right|^2\left|\hat\phi\left(k-\ell'\right)\right|^2
\hat R(\eps^a\ell')dk d\ell'\right\},
$$
which, as $\eps\to0+$, tends to
$$
\hat R(0)\int_{\bbT}\hat R(\ell)d\ell \left\{\int_{\bbR} \left|\hat\phi\left(k\right)\right|^2dk \right\}^2.
$$
Thus, condition \eqref{finite-energy1}  is clearly satisfied by this family of fields.
In the case $a=0$  (microscopically varying initial data) the condition is also valid, as then
$$
\hat\psi^{(\eps)}(k)=\eps^{-1/2}\int_{\bbT}\tilde\phi(k-\ell)\hat \eta(d\ell),
$$
with $\tilde\phi(k)$ given by \eqref{tphi}.
As a result
$$
 \int_\bbT dk\; \left\{\eps\bbE|\hat \psi^{(\eps)}(k)|^2
     \right\}^2= \int_{\bbT}dk \left\{\int_{\bbT} \left|\tilde\phi\left(k-\ell\right)\right|^2\hat R(\ell)d\ell\right\}^2<+\infty.
$$

\subsubsection{Locally stationary initial data}

\label{sec:local-stationary}

Assume that $({\frak r}_{x,\eps},{\frak p}_{x,\eps})_{x\in\bbZ}$, $\eps\in(0,1]$ is a family
locally stationary random fields over a probability space $(\Om,{\cal
  F},\bbP)$. By the above we mean the fields that
satisfy the following:
\begin{itemize}
\item[1)]
they  are square integrable for each $\eps$ and there exist
$C_0^\infty$ functions $r,p:\bbR \to\mathbb R$, called the {\em mean profiles} 
satisfying
$$
 {\frak r}_{x,\eps}'':=\bbE {\frak r}_{x,\eps}=r(\eps x),\qquad  {\frak p}_{x,\eps}'':= \bbE {\frak p}_x^{(\eps)}=p(\eps x),
$$
\item[2)] the
 covariance matrix of the field
$$
{\frak r}_{x,\eps}':= {\frak r}_{x,\eps}- {\frak r}_{x,\eps}'',\quad
{\frak p}_{x,\eps}':= {\frak p}_{x,\eps}- {\frak p}_{x,\eps}'',
$$
is given by
$$
\bbE[ {\frak r}_{x,\eps}'{\frak r}_{x+x',\eps}']=C_{11}(\eps x,x'),\quad \bbE[ {\frak p}_{x,\eps}'{\frak p}_{x+x',\eps}']=C_{22}(\eps x,x') ,
$$ 
$$
\bbE[ {\frak r}_{x,\eps}'{\frak p}_{x+x',\eps}']=C_{12}(\eps
x,x'),\quad \bbE[ {\frak p}_{x,\eps}'{\frak r}_{x+x',\eps}']=C_{21}(\eps x,x') ,\quad x,x'\in\bbZ,
$$ 
where
$C_{ij}:\bbR\times
\bbZ\to\mathbb R$, $i,j=1,2$ are functions that satisfy 
\begin{equation}
\label{011511}
C_*:=\sum_{i=1}^2\sum_{x'}\left|\int_{\bbR}C_{ii}(x,x')dx\right|^2<+\infty.
\end{equation}
\end{itemize}
The energy spectrum of the field   $({\frak r}_{x,\eps}',{\frak
  p}_{x,\eps}')_{x\in\bbZ}$, cf \eqref{011611}, equals
$$
{\frak w}_\eps'(k)=\sum_x C_{22}(\eps x,k)+\frac{\hat
    \al(k)}{4{\frak s}^2(k)} \left(\sum_x C_{11}(\eps x,k)\right)
$$
where
$$
\hat C_{ij}(\eps x,k)=\sum_{x'} C_{ij}(\eps x,x')\exp\left\{-2\pi i
  kx'\right\},\quad (x,k)\in\bbZ\times \bbT,\,i,j=1,2.
$$
We can write
$$
\eps^2 \int_{\bbT}({\frak w}_\eps')^2dk\le C\eps^2\sum_{j=1}^2\sum_{x_1,x_2}
 \int_{\bbT}\hat C_{jj}(\eps x_1,k) \hat C_{jj}^*(\eps x_2,k)dk,\quad \eps\in(0,1],
$$
for some constant $C>0$.
By an applications of  the
Plancherel identity  we conclude that the right hand side approximates
$CC_*$, as $\eps\ll1$. Thus condition 
\eqref{012403} holds in this case.

\subsubsection{Local Gibbs measures}
\label{sec:local-gibbs-measures}

Another important example of  random initial data   is
furnished  by the local Gibbs measure. Given the profiles of temperature
$\beta(\eps x)^{-1}$, momentum $ p(\eps x)$, and
tension $ \tau(\eps x)$, where $\beta^{-1},p,\tau\in
C_0^\infty(\bbR)$ and $\beta^{-1}\ge0$
we define a product measure
analogous to \eqref{eq:gibbs}, in which the constant profiles are replaced by
slowly varying functions. These measures are formally written as
\begin{equation}
  \label{eq:gibbs-loc}
d\nu_{\blambda,\eps} :=  \prod_x \exp\left\{-\beta(\eps x)
  \left(\vphantom{\int_0^1}\gen_x - p(\eps x) \gp_x
    -\tau(\eps x) \gr_x\right) - \mathcal G(\blambda(\eps x))\right\} d\gr_x \; d\gp_x, 
\end{equation}
where $\blambda(x) = \left(\beta(x), p(x), \tau(x) \right)$ {and
$\mathcal G(\cdot)$ is an appropriate Gibbs potential that normalizes the respective measure. In order to make the above ''definition''
rigorous one would have to consider the  Gibbs measure in question as  solutions of the respective DLR equations.  We shall omit that issue
by dealing only with    local Gibbs measures, i.e. the case when $\beta(y)^{-1}$ is compactly supported so the relevant measure is defined on  a finite dimensional space.}
On the sites
where $\beta(\eps x)^{-1} = 0$, we let the corresponding exponential factor in
\eqref{eq:gibbs-loc} be   a delta distribution concentrated at the
point $(0,0)$.  
The corresponding profile of volume stretch $r(\eps x)$ and temperature $\beta(\eps x)^{-1}$ are given by analogues of relations \eqref{tau-n} with $r$
and $u$ appearing there replaced by the respective slowly varying functions. 

The natural decomposition in thermal and mechanical initial conditions
is now given by 
\begin{equation*}
  \gr_x = r(\eps x) + \gr_x' , \qquad \gp_x = p(\eps x) + \gp_x'
\end{equation*}
where $(\gr_x', \gp_x')_x$ are distributed by
\begin{equation}
  \label{eq:gibbs-loc-centered}
d\nu_{\blambda,\eps} ':=    \prod_x \exp\left\{-\beta(\eps x)
  \gen_x' - \mathcal G(\beta(\eps x), 0,
     0)\right\} d\gr_x' \; d\gp_x' \  
\end{equation}
and $\gen_x'$ is the energy at site $x$ of the configuration $(\gr_x',\gp_x')_{x\in\bbZ}$.
In this case we have 
\begin{equation}
\label{032411}
C_{22}(\eps x,x')=\delta_{x',0}\langle
(\gp_x')^2\rangle_{\nu_{\blambda,\eps}'}=\delta_{x',0}\beta^{-1}(\eps x).
\end{equation}

\subsubsection{Nearest neighbor interactions}
Consider first the nearest neighbor case, i.e.  when
$
{\frak e}_{x}'=V(\frak{r}_x').
$
Then,
$$
C_{11}(\eps x,x')=\delta_{x',0}\langle
(\gr_x')^2\rangle_{\nu_{\blambda,\eps}'}=\delta_{x',0}\frac{\int_{\bbR}r^2e^{-\beta(\eps x)V(r)}dr}{\int_{\bbR}e^{-\beta(\eps x)V(r)}dr}.
$$
Let us assume that there exists $c_*>0$, for which $c_*r^2\le V(r)$. Then,
\begin{eqnarray*}
&&
0\le C_{11}(\eps x,x')\le
\frac{1}{c_*}\delta_{x',0}\frac{\int_{\bbR}V(r)e^{-\beta(\eps
    x)V(r)}dr}{\int_{\bbR}e^{-\beta(\eps x)V(r)}dr}\\
&&
=\frac{1}{c_*}\delta_{x',0} \left(\tilde u(\eps x)-\frac12\beta^{-1}(\eps x)\right).
\end{eqnarray*}
Here $\tilde u(\eps x) := u(0,\beta(\eps x))$ and $u(\tau,\beta)$ is the internal energy function defined in
\eqref{tau-n}, see e.g. (2.1.9) of \cite{BO}. Condition \eqref{011511}
is satisfied, once we assume that
$$
\int_{\bbR}\beta^{-1}(y)dy<+\infty\quad\mbox{and}\quad
\int_{\bbR}\tilde u(y)dy<+\infty.
$$


\subsubsection{Gaussian, locally Gibbs measures}

\label{lgg}

We assume that ${\rm supp}\,\beta^{-1}= [-K,K]$ and that $\beta^{-1}(y)>0$ for $|y|<K$. In addition, we suppose that
the sequence $(\al_x)$, besides satisfying conditions a1)-a3) and \eqref{al-0}, is compactly supported, i.e. there exists a positive integer $\ell$ such that $\al_x=0$ for $|x|>\ell$ and $\al_{x}<0$ for $0<|x|\le \ell$ (obviously in light of \eqref{al-0} we have $\al_0>0$).
In this case, the formal expression \eqref{eq:gibbs-loc-centered} 
is a probability measure on $\ell_2(\bbZ)$ that equals (we omit writing primes)
\begin{equation}
  \label{eq:gibbs-loc-centered1}
\nu_{\blambda,\eps}:= \mu_{\eps,K}  \otimes  \delta_{K^c},   
\end{equation}
where $\delta_{K^c}$ is the Borel probability measure on the space of sequences $({\frak r}_x)_{|x|\ge \eps^{-1}K}$, concentrated on the sequence
$ {\frak r}_x\equiv 0$, $|x|\ge K/\eps$. Measure  $\mu_{\eps,K}$ is  Gaussian on the Euclidean space
corresponding to finite sequences $(\rho_x)_{|x|< \eps^{-1}K}$ whose characteristic functional equals
$$
\exp\left\{-\frac12\sum_{|x|,|x'|<\eps^{-1}K}\sigma_{x,x'}\rho_x\rho_{x'}\right\},
$$
where $\Sigma:=[\sigma_{x,y}]$ is the inverse of the symmetric  matrix $S:=[S_{x,y}]$  corresponding to the quadratic form
$$
\sum_{|x|,|x'|<\eps^{-1}K}S_{x,x'}{\frak r}_x{\frak r}_{x'}=-\frac{1}{4}\sum_{|x|,|x'|<\eps^{-1}K}\beta(\eps x)\al_{x-x'}{\frak q}_{x,x'}^2,
$$
where ${\frak q}_{x,x'}$ is defined in \eqref{diff-q}.
 We claim that there exists $c_*>0$
independent of $\eps$ and such that
\begin{equation}
\label{012411}
\sum_{|x|,|x'|<\eps^{-1}K}S_{x,x'}{\frak r}_x{\frak r}_{x'}\ge c_*\sum_{|x|<\eps^{-1}K}{\frak r}_x^2,\quad \forall\,({\frak r}_x)_{|x|<\eps^{-1}K}.
\end{equation}
According to our assumptions there exists $\beta_*>0$ such that $\beta(y)\ge \beta_*$ for all $y\in\bbR$.
We can write therefore
\begin{eqnarray}
\label{022411}
&&
\inf\left\{\sum_{|x|,|x'|<\eps^{-1}K}S_{x,x'}{\frak r}_x{\frak r}_{x'}:\,\sum_{|x|<\eps^{-1}K}{\frak r}_x^2=1\right\}\\
&&
\ge \frac{\beta_*}{4}\inf\left\{-\sum_{|x|,|x'|<\eps^{-1}K}\al_{x-x'}{\frak q}_{x,x'}^2:\,\sum_{|x|<\eps^{-1}K}{\frak r}_x^2=1\right\}\nonumber\\
&&
\ge  \frac{\beta_*}{4}\inf\left\{-\sum_{x,x'}\al_{x-x'}{\frak q}_{x,x'}^2:\,\sum_{x}{\frak r}_x^2=1\right\}.\nonumber
\end{eqnarray}
One can easily verify the following identity (cf \eqref{om2bb})
$$
-\frac{1}{2}\sum_{x,x'}\al_{x-x'}{\frak q}_{x,x'}^2=\tau_1\int_{\bbT}\varphi({\frak s}^2(k))|\hat{\frak r}(k)|^2dk,
$$
therefore the utmost left hand side \eqref{022411} can be estimated from below by
$$
 c_*:=\frac{\tau_1\beta_*}{2}\inf\varphi.
$$
Hence \eqref{012411} holds with $c_*$ as defined above.

We can write
$$
\langle |\hat {\frak r}(k)|^2\rangle_{\nu_{\blambda,\eps}}=
\sum_{|x|,|x'|<\eps^{-1}K}\si_{x,x'}e_x(k)e_{x'}^*(k),
$$
with $e_x(k):=\exp\left\{-2\pi i x k\right\}$. Thanks to \eqref{012411} we obtain
\begin{equation}
\label{052411}
\langle |\hat {\frak r}(k)|^2\rangle_{\nu_{\blambda,\eps}}\le\frac{1}{c_*}
\sum_{|x|\le \eps^{-1}K}|e_x(k)|^2=\frac{2(K+1)}{c_*\eps},\quad \forall\,k\in\bbT.
\end{equation}
Thanks to  \eqref{032411} we conclude that there exists $C>0$ such that
\begin{equation}
\label{062411}
\langle |\hat {\frak p}(k)|^2\rangle_{\nu_{\blambda,\eps}}=\sum_{x}\langle {\frak p}^2_x\rangle_{\nu_{\blambda,\eps}}
=\sum_{x}\beta^{-1}(\eps x)\le \frac{C}{\eps},\quad \forall\,\eps\in(0,1].
\end{equation}
Combining \eqref{052411} with \eqref{062411} we conclude condition \eqref{finite-energy1}.
%

\appendix

\section{Proof of Proposition \ref{prop011706}}

\label{appA}

A simple calculation, using the fact that $\sum_x\al_x=0$, shows that
\begin{equation}
\label{011206}
\sum_x(\al*{\frak q})_x{\frak q}_x=-\frac{1}{2}\sum_{x,y}\al_{x-y}({\frak q}_x-{\frak q}_y)^2
\end{equation}
for any $({\frak q}_x)$ such that ${\frak r}:=\nabla^*{\frak q}$
is square summable.

Using \eqref{011206}, the Plancherel identity and \eqref{om2bb} we can write
\begin{eqnarray}
 &&
 \sum_x {\frak e}_x ({\frak r}, {\frak p})=\int_{\bbT}\left(\frac{|\hat{\frak p}(k)|^2}{2}+\hat\al(k)|\hat{\frak q}(k)|^2\right)dk\nonumber\\
 &&
=\int_{\bbT}\left(\frac{|\hat{\frak p}(k)|^2}{2}+\tau_1\varphi^2({\frak s}^2( k))|\hat{\frak r}(k)|^2\right)dk.
\end{eqnarray}
Here $\hat{\frak r}(k)$, $\hat{\frak p}(k)$, $\hat{\frak q}(k)$ are the Fourier transforms of 
${\frak r}_x$, ${\frak p}_x$, $\hat{\frak q}_x$ respectively. In light of the assumptions made about $\varphi(\cdot)$ it is clear that 
the utmost right hand side is equivalent to $\sum_{x}\left({\frak r}_x^2+ {\frak
  p}_x^2\right)$, so \eqref{equiv1} follows.

Obviously only the lower bound in \eqref{equiv} requires a proof.  Note that 
 \begin{equation}
 \label{021706}
 \sum_x|{\frak
   e}_{x}|\le  \frac{1}{2}\sum_x{\frak
   p}_{x}^2+
  \sum_{x,x'} |\alpha_{x-x'}|{\frak q}_{x,x'}^2.
 \end{equation}
 Here ${\frak q}_{x,x'}$ is given by \eqref{diff-q}.
   By Cauchy-Schwartz inequality for $x\ge x'$ we have
 $$
 {\frak
   q}_{x,x'}^2\le (x-x')\sum_{x'<x''\le x} {\frak r}_{x''}^2
 $$
 and an analogous inequality holds also for  $x< x'$. Therefore
 \begin{eqnarray*}
 &&\sum_{x,x'}|\al_{x-x'}|
 {\frak q}_{x,x'}^2 \le \sum_{x\ge x''> x'}|x-x'||\al_{x-x'}|
 {\frak r}^2_{x''}\\
 &&
 +\sum_{x'\ge x''> x}|x-x'||\al_{x-x'}|
 {\frak r}^2_{x''}.
 \end{eqnarray*} 
 Substituting $z:=x-x'$ in the first summation and $z:=x'-x$ in the
 second we get
 that the right hand side equals
 $$
 \sum_{z>0}z|\al_{z}|\sum_{x\ge x''> x-z}
 {\frak r}^2_{x''}
 +\sum_{z>0}z|\al_{z}|\sum_{x+z\ge x''> x}
 {\frak r}^2_{x''}.
 $$
 Denote
 the first and the second term by $I$ and $I\!I$ respectively.
 We have
 \begin{eqnarray*}
 &&
 I= \sum_{z>0}z|\al_{z}|\sum_{x''}  {\frak r}^2_{x''}\sum_{x''+z>x\ge x''}1\\
 &&
 \le \left(\sum_{z>0}z^2|\al_{z}|\right)\sum_{x''} {\frak
     r}^2_{x''}.
 \end{eqnarray*} 
 On the other hand 
 \begin{eqnarray*}
 &&
 I\!I= \sum_{z>0}z|\al_{z}|\sum_{x''} 
{\frak
     r}^2_{x''}\sum_{x+z\ge  x''>x}1
 \le \left(\sum_{z>0}z^2|\al_{z}|\right)\sum_{x''} {\frak
     r}^2_{x''}
      \end{eqnarray*} 
and, as a result, we get
\begin{equation}
 \label{021706a}
 \sum_x|{\frak
   e}_{x}|\le  C\sum_x({\frak
   p}_{x}^2+{\frak
   r}_{x}^2),
 \end{equation}
 where
 $C:=\max\left\{1/2,\sum_{z>0}z^2|\al_z|\right\}$.
This combined with the already shown estimate \eqref{equiv1} ends the proof of the lower bound in \eqref{equiv}.


\end{document}